\documentclass[preprint,nofootinbib]{revtex4}%
\usepackage{amssymb}
\usepackage{amsfonts}
\usepackage{amsmath}
\usepackage{amsmath}
\usepackage{graphicx}
\usepackage{hyperref}
\usepackage[usenames]{color}%
\providecommand{\U}[1]{\protect\rule{.1in}{.1in}}
\hypersetup{colorlinks,linkcolor={blue},citecolor={blue},urlcolor={black}} 
\providecommand{\U}[1]{\protect\rule{.1in}{.1in}}
\definecolor{blue}{rgb}{0,0,1}

\definecolor{red}{rgb}{1,0,0}

\begin{document}

\title{More on near-horizon charges black holes with gravitational hair in three dimensions}

\author{Seyed Naseh Sajadi}
\email{naseh.sajadi@gmail.com}
\affiliation{Strong Gravity Group, Department of Physics, Faculty of Science, Silpakorn University, Nakhon Pathom 73000, Thailand}

\author{Supakchai Ponglertsakul}
\email{supakchai.p@gmail.com}
\affiliation{Strong Gravity Group, Department of Physics, Faculty of Science, Silpakorn University, Nakhon Pathom 73000, Thailand}

\author{Julio Oliva}
\email{julioolivazapata@gmail.com}
\affiliation{Departamento de Física, Universidad de Concepción, Casilla, Concepcion, 160-C Chile}

\begin{abstract}
With the aim of continuing the exploration of near-horizon charges in higher-curvature gravity, {searching for sectors leading to universal behaviors}, we first provide a thorough revision and formulae of the covariant phase-space method applied to arbitrary gravitational theories containing up to quartic terms in the Riemann tensor in arbitrary dimension. {These results can be applied in diverse setups, in particular in the context of $\alpha'$ corrections to String Theory, where it is known that in Type II theories, the first correction to the Einstein-Hilbert Lagrangian goes as $\alpha'^3 \mathcal{R}^4$}. Then, we test these formulae for near horizon asymptotic symmetries of the rotating BTZ spacetime where the first law of black hole thermodynamics is consistently recovered. It was recently realized that a subset of these higher curvature gravities do admit black holes with gravitational hair, whose entropy can be microscopically accounted for, as is the case of New Massive Gravity. In this case, the four maximally symmetric vacua of the theory coincide, and the theory acquires an extra gauge symmetry when linearized around such a vacuum. We study the near-horizon asymptotic symmetries and compute the associated charges, both in the static and rotating hairy black holes, extending up to $\mathcal{R}^4$, a work that was previously done only up to a quadratic term. {In order to allow for a continuous lecture on the work, we report the explicit expressions of the general Lagrangians in the appendices.} 
\end{abstract}

\maketitle

\section{Introduction}
Since Emmy Noether’s seminal theorems, conserved charges have intrinsically been tied to the notion of symmetry. In General Relativity (GR), however, the generally covariant nature of the theory and the absence of a globally defined time direction make the covariant definition of conserved charges highly nontrivial. From the early days of GR, numerous proposals have been put forward to compute conserved quantities associated with exact space-time isometries, via Killing vector fields, or with asymptotic symmetries in flat and anti–de Sitter (AdS) spacetimes. More than a century later, this field has developed into a rich and well-established body of literature, though important challenges remain. In particular, a consistent formulation of local conserved charges is still lacking, while there are several successful approaches to defining quasi-local conserved charges \cite{Szabados:2004xxa, Banados:2016zim}. Among the different approaches, two main lines of formulation can be distinguished: one is the Hamiltonian formulation, which is based on space and time decomposition, and the other is the Lagrangian formulation, which is based on spacetime covariance. 
 
A precursor of the Hamiltonian formulation is the introduction of quasi-local charges by Komar. In Komar's method, the quasilocal mass and angular momentum for asymptotically flat solutions can be found by integrating over a codimension-2 surface in constant-time at infinity. Soon after, the Hamiltonian formulation of gravitational theories was elaborated in a series of works from 1959 to 1962 by Arnowitt, Deser, and Misner, known as the ADM formulation \cite{Arnowitt:1962hi}. After that, a similar formulation for null asymptotics is proposed by Bondi et al. in 1962 \cite{Bondi:1962px}. The Hamiltonian formulation for asymptotic flat spacetimes reached its mature formalism by Regge-Teitelboim in 1974, emphasizing the role of surface terms in the Hamiltonian formulation \cite{Regge:1974zd}.
Nevertheless, there is a shortcoming of the formulation when the asymptotic flatness is relaxed to include asymptotic AdS solutions, mainly because of the appearance of divergent conserved charges. {This problem, interesting in its own merit in order to understand all the healthy sectors of the theory, became particularly relevant in the context of gauged supergravities, where AdS, BPS backgrounds, naturally appear due to the charged nature of the gravity (see \cite{Trigiante:2016mnt} and references therein)}. Later progress in this line of formulation has been mainly in the direction of solving this problem \cite{Hollands:2005wt}. 

The Hamilton-Jacobi approach was formulated in the seminal paper by Brown and York in 1992 \cite{Brown:1992br} where later in \cite{Balasubramanian:1999re}, a surface counterterm to the Lagrangian was added.  Reformulation of conserved charges based on covariantly defined conserved currents and subtracting the contributions from a reference solution has been proposed as another method by Abbott-Deser in 1982 \cite{Abbott:1981ff}. This approach is later extended to higher curvature theories by Deser-Tekin \cite{Deser:2002rt,Deser:2002jk}.

The second approach to defining conserved charges is based on the Lagrangian formulation, which inherits the covariant structure by construction. This method was firstly introduced by Ashtekar and Crnkovic–Witten in 1987 \cite{Ashtekar:1987hia,Crnkovic:1986ex} and later was systematically developed by various authors \cite{Lee:1990nz,Wald:1993nt,Iyer:1994ys}.  In this framework, called the covariant phase-space formulation, the phase space is constructed directly from the dynamical fields across spacetime, without spacetime decomposition. The symplectic form is derived from the Lagrangian, which provides a precise definition of conserved charges associated with diffeomorphisms and gauge transformations. A key result of this formulation is that black hole entropy can be understood as a conserved charge evaluated at the horizon \cite{Wald:1993nt}.

In this paper, we apply the covariant phase-space method to higher-curvature gravity theories, with a particular focus on cubic and quartic terms built from the Riemann and Ricci tensors. {Terms up to this order appear in different perturbative formulations of quantum gravity, as for example in Type II String Theories where the first stringy, $\alpha'$ correction to the Einstein-Hilbert action, emerges at order $\alpha'^3$ which multiplies an eight-derivatives term in the action of order $\mathcal{R}^4$ which denotes a precise (up to field redefinitions), quartic combination of the Riemann curvature \cite{Gross:1986iv,Grisaru:1986px}. On the other hand, it is also well-known that perturbative quantum gravity in vacuum requires a single counterterm at two-loops. Such counterterm, dubbed Goroff-Sagnotti term is cubic in the Riemann tensor \cite{Goroff:1985sz} (the physical effects at the level of black hole theory for these types of terms have been explored in \cite{Daas:2023axu} (see also references therein). 

These two frameworks define clear potential applications of the formulae which we will obtain in what follows. Due to their intricate explicit form, they will be relegated to the appendices. We begin with a brief review of the solution phase-space method, followed by its application to higher-curvature theories up to terms that are quartic in the Riemann tensor, in full generality and in arbitrary dimensions. Then we move to $2+1$-dimensions, where higher curvature terms can be introduced with non-perturbative couplings in a healthy manner, as reviewed below, leaving the applications mentioned above in String Theory and pertubative quantum gravity for future studies. We illustrate the framework through several representative examples in three-dimensional higher curvature gravity, focusing on the theories that admit black holes with hair of gravitational origin, both in the static and rotating cases, both at infinity and in the near horizon regions.}

\section{Covariant phase space}

The covariant phase-space method provides a systematic way of calculating variations or perturbations of conserved charges in generic theories with local symmetries and generally in covariant theories. This method is primarily developed in the papers by Wald et al. \cite{Lee:1990nz, Wald:1993nt, Iyer:1994ys}. Wald's approach, which we will succinctly review in the following, is based on the Lagrangian action formulation. A modern way of formulating the covariant phase-space method based on dynamics is further developed by Barnich and Brandt in \cite{Barnich:2001jy}.  A concrete discussion can be found in the original papers, while for a pedagogical review, we refer the interested readers to  \cite{Hajian:2015eha,Ghodrati:2016vvf}. 

The phase space is a manifold that is equipped with a symplectic two-form $\Omega$. To construct $\Omega$ for a given $d$-dimensional generally invariant theory, one may
start by writing the action
\begin{equation}
    S[\phi]=\int L,
\end{equation}
where $L$ is the Lagrangian $d$-form and $\phi$ is used to denote all dynamical fields. The variation of the action to the fields $\phi$ is given by
\begin{equation}
\delta L[\phi]=E_{\phi}\delta\phi+d\Theta(\phi,\delta \phi),
\end{equation}
where $\delta\phi$ is a generic field variation and $E_{\phi}=0$ denotes the field equations for the field $\phi$, and $\Theta$ is the symplectic potential picked up from the surface term of the variation. The ($d-1$)-form symplectic current is $\omega=\delta_{1}\Theta(\phi,\delta_{2}\phi)-\delta_{2}\Theta(\phi,\delta_{1}\phi)$ and the symplectic form over a Cauchy surface is 
\begin{equation}
\Omega(\delta_{1}\phi,\delta_{2}\phi,\phi)=\int_{\Sigma}\omega(\delta_{1}\phi,\delta_{2}\phi,\phi),
\end{equation}
where $\delta_{1}\phi$ and $\delta_{2}\phi$ are two arbitrary field perturbations and $\Sigma$ is a codimension$-1$ spacelike surface. There are two kinds of ambiguities present in the covariant phase space formulation. The first comes from the fact that the formulation is based on the Lagrangian formulation, which is itself ambiguous up to a surface term $L\to L+d\mathcal{K}$. It results in $\Theta\to\Theta+\delta\mathcal{K}$, but $\omega$ remains intact. A second ambiguity originates from the definition of $\Theta$, one can add an exact $(d-1)$-form
$d\mathcal{Y}$ to $\Theta$, 
\begin{equation}
    \Theta\to\Theta=\Theta+d\mathcal{Y},
\end{equation}
As a result, the symplectic current inherits the following transformation 
\begin{equation}    \omega(\delta_{1}\phi,\delta_{2}\phi,\phi)\to\omega(\delta_{1}\phi,\delta_{2}\phi,\phi)+d(\delta_{1}\mathcal{Y}(\delta_{2}\phi,\phi)-\delta_{2}\mathcal{Y}(\delta_{1}\phi,\phi)).
\end{equation}
We may use $\omega$ to define conserved charges associated with a specific set of transformations generated by a Killing vector $\xi$. Since $d\omega(\delta_{1}\phi,\delta_{2}\phi,\phi)=0$ for any two perturbations satisfying linearized equations of motion, we have
\begin{equation}
    \omega_{\xi}(\delta\phi,\delta_{\xi}\phi,\phi)=dk_{\xi}(\delta\phi,\phi),
\end{equation}
where $k_{\xi}$ is a $(d-2)-$form. We may integrate $k_{\xi}$ to define the perturbations of the charge associated with $\xi$
\begin{equation}\label{eqcharge}
\delta H_{\xi}=\oint_{\partial \Sigma}k_{\xi}(\delta\phi,\phi).
\end{equation}
 The $(d-2)-$form $k_{\xi}$ can be shown explicitly to be 
\begin{equation}\label{eqkxi}
k_{\xi}=\delta Q_{\xi}-\xi.\Theta,
\end{equation}
in which $Q_{\xi}$ is the Noether-Wald charge density. It is  defined by the relation
 \begin{equation}
dQ_{\xi}=\Theta-\xi .L.
\end{equation}
By obtaining $k_{\xi}$ from \eqref{eqkxi}, one can evaluate the charge using \eqref{eqcharge}. If the result were integrable, then an integration over $\delta\phi$ would produce a finite charge.
Now let us consider a stationary black hole solution with a Killing field $\xi$ which generates a Killing horizon and vanishes on a bifurcation surface $\mathcal{H}$. If we choose the hypersurface $\Sigma$ to have its outer boundary at spatial infinity and the interior boundary at $\mathcal{H}$, then the variational identity can be expressed with two boundary terms
\begin{equation}\label{eqtermo}
   \int_{H} k_{\xi} =\int_{\infty} k_{\xi},
\end{equation}
for $\delta_{\xi}\phi=0$. If we assume that the asymptotic symmetries have been specified by the time translational Killing field and axial rotational one with the horizon angular velocity $\Omega_{H}$, i.e., $\xi=\partial_{t}+\Omega_{H}\partial_{\phi}$. Then the outer boundary integral of \eqref{eqtermo} can be defined as the total energy and the angular momentum.
Comparing \eqref{eqtermo} with the first law of thermodynamics $T\delta S=\delta M-\Omega_{H}\delta J$, the left-hand side gives the black hole entropy in the form
\begin{equation}\label{eqeqntropydef}
    S=\dfrac{2\pi}{\kappa}\int_{\mathcal{H}}Q_{\xi}.
\end{equation}
Here $\kappa$ is the surface gravity of the unperturbed black hole. The variation of the mass and angular momentum has the form
\begin{equation}\label{eqqmasang}
    \delta M=\dfrac{1}{16\pi G}\int_{\infty}k_{\xi}[\partial_{t}],\;\;\;\;\; \delta J=\dfrac{1}{16\pi G}\int_{\infty}k_{\xi}[\partial_{\phi}].
\end{equation}
After setting up the generic formulation, we apply this method to cubic and quartic gravity in the following sections. {At this point, we emphasize that the purpose of introducing the general higher-curvature Lagrangian in \eqref{eqacti} and \eqref{lagquart} is not to analyze every possible theory individually, but to establish a unified and theory-independent framework for constructing conserved charges in $d-$dimensional gravity, with the further motivation of particular relevant scenarios developed in the introduction.}

\section{Einstein gravity supplemented by cubic terms in the curvature}
In this section, we deal with the Einstein-Hilbert action, augmented by the most general combination of terms which are cubic in the Riemann tensor, and compute the conserved current provided by the covariant phase space formalism. Einstein gravity coupled to the mentioned cubic terms in the Riemann tensor can be parameterized as follows
\begin{align}\label{eqacti}
I=\dfrac{1}{2\kappa}\int &d^{d}x \sqrt{-g}\Big\lbrace  R-2\Lambda+\alpha_{0} f(R)+\alpha_{1} R_{a}^{c}R_{c}^{b}R_{b}^{a}+\alpha_{2} RR_{ab}R^{ab}+\alpha_{3} R_{a b}{}^{c d}R_{c d}{}^{e f}R_{e f}{}^{a b}+\nonumber\\
&\alpha_{4} R_{a b c d}R^{a c}R^{b d}+\alpha_{5} R R^{abcd}R_{abcd}+\alpha_{6} R^{abcd}R_{cdbe}R^{e}_{a}+\alpha_{7} R_{a}{}^{c}{}_{b}{}^{d}R_{c}{}^{e}{}_{d}{}^{f}R_{e}{}^{a}{}_{f}{}^{b}
\Big\rbrace ,
\end{align}
where $R^{a}{}_{bcd}$, $R_{ab}$, and $R$ are the Riemann tensor, the Ricci tensor, and the Ricci scalar, respectively. Before we continue, it should be mentioned that the calculations for the charges of the Einstein-quadratic gravity using the covariant phase space method are presented in \cite{Ghodrati:2016vvf} and using the Abbott-Deser-Tekin (ADT) method in \cite{Devecioglu:2010sf, Nam:2010ub}. In this section, we focus only on the cubic term and report the main results. Detailed analyses are similar to the simple Einstein-Hilbert Lagrangian, which can be found in \cite{Hajian:2015eha}. {Notice, as well, that in arbitrary dimensions, there are eight independent, algebraic invariants, containing three Riemann tensors \cite{Fulling:1992vm}. Here, we conveniently parameterized the eight independent terms via the eight couplings $\alpha_i$ with $i=0,\ldots,7$. The term $R^3$ is contained in the general $f(R)$ term which is multiplied by the coupling $\alpha_0$ in \eqref{eqacti}.}

By variation of Lagrangian \eqref{eqacti}, the equation of motion is given by, 
\begin{equation}
\mathcal{E}_{mn}=\mathcal{E}^{Ein}_{mn}+\sum_{i=0}^{\text{7}}\mathcal{E}^{\alpha_{i}}_{mn}.
\end{equation}
The detailed expression of these terms is given in Appendix \ref{fileq} (see also Appendix B of \cite{Oliva:2010zd}). {Notice that the whole structure of the equations is controlled by the four-rank tensor $\mathcal{P}^{abcd}=\frac{\partial\mathcal{L}}{\partial R_{a b c d}}$, as mentioned in the Appendix \ref{fileq}. Due to their intricate explicit structure, we have performed several symmetry cross checks of these expressions, which could be also performed with the xCPS Package \cite{xCPS1,xCPS2} on the xAct bundle \cite{xAct}. At the level of cubic invariants, the expressions are not very long and do not interrupt the flow or readability of the manuscript. Therefore, below we report the Sympletic Potential, the Noether-Wald Charge for the general combination of cubic invariant and mention how the trivial combination of the variation of the latter (in field space), the former and a vector field, permit constructing the Surface charges. For the quartic theory, we will heavily refer to Appendix \ref{expressionsquartic}.
}

\paragraph{Symplectic potential} By variation of the Lagrangian and using the equation of motion (EoM), the surface $(d-1)$-form $\Theta$ can be read as
\begin{equation}
    \Theta^{m}=\sqrt{-g}\Theta^{m}_{Ein}+\sqrt{-g}\sum_{i=0}^{\text{7}}\Theta^{m}_{\alpha_{i}},
\end{equation}
with 
{\small
\begin{align}\label{fofRTheta}
\Theta^{m}_{\alpha_{0}}&=f^{\prime}\left(\nabla_{a}h^{a m}-\nabla^{m}h\right)-\nabla_{a}f^{\prime}h^{m a}+
\nabla^{m}f^{\prime}h,\\
\Theta^{m}_{\alpha_{1}}&=3\nabla_{c}h^{m}_{b}R^{a b}R_{a}^{c}-3h_{b}^{d}\nabla_{d}\left(R^{ab}R_{a}^{m}\right)-\dfrac{3}{2}\nabla^{m}h_{bc}R^{ab}R_{a}^{c}+\dfrac{3}{2}h_{bc}\nabla^{m}\left(R^{ab}R_{a}^{c}\right)-\dfrac{3}{2}\nabla_{b}hR_{a}^{m}R^{ab}\nonumber\\
&+\dfrac{3}{2}h\nabla_{c}\left(R_{a}^{c}R^{am}\right),\\
\Theta^{m}_{\alpha_{2}}&=-\nabla_{a}hRR^{am}+h\nabla_{b}\left(RR^{m b}\right)+2\nabla_{b}h_{a}^{m}RR^{ab}-2h_{a}^{c}\nabla_{c}\left(RR^{am}\right)- \nabla^{m}h_{ab}RR^{ab}\nonumber\\
&+h_{ab}\nabla^{m}\left(RR^{ab}\right)+\nabla_{c}h^{cm}
R_{ab}R^{ab}-h^{m d}\nabla_{d}\left(R_{ab}R^{ab}\right)-\nabla^{m}hR_{ab}R^{ab}+
h\nabla^{m}\left(R_{ab}R^{ab}\right),\\
\Theta^{m}_{\alpha_{3}}&=-6R_{ac}{}^{ef}R_{b}{}^{m}{}_{ef}\nabla^{c}h^{ab}+6h^{ab}\nabla^{d}\left(R_{a}{}^{m e f}R_{b d e f}\right),\\
\Theta^{m}_{\alpha_{4}}&=\nabla_{b}h_{ac}R^{cm}R^{ab}-h_{ac}\nabla_{d}\left(R^{cd}R^{am}\right)-\nabla_{c}h_{ab}R^{ab}R^{cm}+
h_{ab}\nabla_{d}\left(R^{ab}R^{m d}\right)-\nabla^{m}h^{cd}R^{ab}R_{acbd}\nonumber\\
&+
h^{cd}\nabla^{m}\left(R^{ab}R_{acbd}\right)+2\nabla_{c}h^{cd}R^{ab}R_{adb}{}^{m}-2h^{m d}\nabla^{e}\left(R^{ab}R_{adbe}\right)-\nabla^{d}hR^{ab}R_{adb}{}^{m}\nonumber\\
&+h\nabla^{e}\left(R^{ab}R_{a}{}^{m}{}_{be}\right),\\
\Theta^{m}_{\alpha_{5}}&=\nabla_{a}h^{am}R_{cdef}R^{cdef}-h^{am}\nabla_{a}\left(R_{cdef}R^{cdef}\right)-\nabla^{m}h R_{cdef}R^{cdef}+h\nabla^{m}\left(R_{cdef}R^{cdef}\right)\nonumber\\
&-4RR_{acb}{}^{m}\nabla^{c}h^{ab}+4h^{ab}\nabla^{d}\left(RR_{a}{}^{m}{}_{bd}\right),
\end{align}
\begin{align}
\Theta^{m}_{\alpha_{6}}&=\dfrac{1}{2}\nabla^{m}h^{ab}R_{a}{}^{def}R_{bdef}-\dfrac{1}{2}h^{ab}\nabla^{m}\left(R_{a}{}^{def}R_{bdef}\right)-\nabla_{a}h^{ab}R_{b}{}^{def}R^{m}{}_{def}+h^{m b}\nabla_{c}\left(R_{b}{}^{def}R^{c}{}_{def}\right)\nonumber\\
&+\dfrac{1}{2}\nabla^{b}hR_{b}{}^{def}R^{m}{}_{def}-\dfrac{1}{2}h\nabla^{c}\left(R^{m def}R_{cdef}\right)-2\nabla_{a}h^{cd}R^{ab}R_{bcd}{}^{m}+2h^{cd}\nabla^{e}\left(R^{m b}R_{bcde}\right)\nonumber\\
&+\nabla^{d}h_{a}^{c}R^{ab}\left(R_{bcd}{}^{m}+R_{bdc}{}^{m}+R_{b}{}^{m}{}_{cd}\right)-h_{a}^{c}\nabla^{e}\left[R^{ab}\left(R_{bc}{}^{m}{}_{e}+R_{b}{}^{m}{}_{ce}+R_{bec}{}^{m}\right)\right],\\
\Theta^{m}_{\alpha_{7}}&=3\nabla^{c}h^{ab}\left(R_{a}{}^{e}{}_{c}{}^{f}R_{b}{}_{f}{}^{m}{}_{e}-R_{a}{}^{e}{}_{b}{}^{f}R_{ce}{}^{m}{}_{f}\right)-3h^{ab}\nabla^{d}\left(R_{a}{}^{e}{}_{b}{}^{f}R^{m}{}_{edf}-R_{a}{}^{e m f}R_{bfde}\right),
\end{align}
}
where $f'=df/dR,h_{mn}=\delta g_{mn}$, $h^{mn}=g^{ma}g^{nb}h_{ab}$ and $h=g^{mn}\delta g_{mn}$.
\paragraph{Noether-Wald charge.}
Having $\Theta$ at hand, by imposing the field equations, the Noether-Wald $(d-2)$-form $Q_{\xi}$ can be read as
\begin{equation}
    \mathcal{Q}^{mn}=\sqrt{-g}Q^{mn}_{Ein}+\sqrt{-g}\sum_{i=0}^{\text{7}} Q^{mn}_{\alpha_{i}},
\end{equation}
with
{\small
\begin{align}\label{NWalpha0}
Q^{mn}_{\alpha_{0}} &= 4\nabla^{[m}f^{\prime}\xi^{n]}-2f^{\prime}\nabla^{[m}\xi^{n]},\\
Q^{mn}_{\alpha_{1}}&=6\nabla^{b}\xi^{[m}R_{a}^{n]}R_{b}^{a}+6\xi^{[n}\nabla_{c}
\left(R^{m]a}R_{a}^{c}\right)+6\xi^{b}\nabla^{[m}\left(R^{n]a}R_{ab}\right),\\
Q^{mn}_{\alpha_{2}}&=4R\nabla^{a}\xi^{[m}R^{n]}_{a}+
2R^{ab}R_{ab}\nabla^{[n}\xi^{m]}+4\xi^{[n}\nabla_{b}\left(R^{m]b}R\right)+
4\xi^{[n}\nabla^{m]}\left(R_{ab}R^{ab}\right)\nonumber\\
&+4\xi^{b}\nabla^{[m}\left(R^{n]}_{b}R\right),\\
Q^{mn}_{\alpha_{3}}&=12\xi^{b}\nabla^{d}\left(R^{[nm]e f}R_{b d e f}\right)-12R_{a}{}^{[n e f}R_{b}{}^{m]}{}_{e f}\nabla^{a}\xi^{b},\\
Q^{mn}_{\alpha_{4}}&=4\xi_{c}\nabla^{[m}\left(R_{a}{}^{n]}{}_{b}{}^{c}R^{ab}\right)
+4\xi^{[n}\nabla^{e}\left(R_{a}{}^{m]}{}_{be}R^{ab}\right)+4\nabla^{[n}\xi^{d}R_{adb}{}^{m]}R^{ab}+2\nabla_{a}\xi_{c}R^{c[m}R^{n]a}\nonumber\\
&+4\xi_{c}\nabla_{d}\left(R^{c[n}R^{m]d}\right),\\
Q^{mn}_{\alpha_{5}}&=2R_{cdef}R^{cdef}\nabla^{[n}\xi^{m]}+4\xi^{[n}
\nabla^{m]}\left(R_{cdef}R^{cdef}\right)+8\xi^{a}\nabla^{c}\left(RR_{ac}{}^{[nm]}\right)-8RR_{b}{}^{[m}{}_{a}{}^{n]}\nabla^{b}\xi^{a},\\
Q^{mn}_{\alpha_{6}}&=2\nabla^{[m}\xi^{b}R^{n]}{}_{def}R_{b}{}^{def}+2\xi^{b}\nabla^{[n}\left(R^{m]}{}_{def}R_{b}{}^{def}\right)+2\xi^{[m}\nabla^{c}\left(R^{n]def}R_{cdef}\right)+2\nabla_{a}\xi^{d}R^{ab}R_{b}{}^{[m}{}_{d}{}^{n]}\nonumber\\
&+2\nabla^{d}\xi_{a}R^{ab}R_{b}{}^{[n}{}_{d}{}^{m]}-\nabla_{a}\xi^{c}R^{ab}R_{bc}{}^{nm}+\nabla^{c}\xi_{a}R^{ab}R_{bc}{}^{nm}-2\xi_{a}\nabla^{d}\left(R^{ab}R_{bd}{}^{nm}\right)\nonumber\\
&+4\xi^{a}\nabla^{d}\left(R^{b[n}R_{ba}{}^{dm]}\right)+4\xi^{a}\nabla^{d}\left(R^{b[m}R_{b}{}^{n]}{}_{ad}\right), \\
Q^{mn}_{\alpha_{7}}&=6\nabla^{b}\xi^{a}R^{[n e}{}_{b}{}^{f}R_{a f}{}^{m]}{}_{e}+6\xi^{a}\nabla^{d}\left(R_{a}{}^{e}{}_{d}{}^{f}R^{[n}{}_{e}{}^{m]}{}_{f}\right)+6\xi^{a}\nabla^{d}\left(R_{a}{}^{e[n f}R^{m]}{}_{edf}\right)\nonumber\\
&+6\xi^{a}\nabla^{d}\left(R^{[n f}{}_{a}{}^{e}R^{m]}{}_{fde}\right).
\end{align}
}
\paragraph{Surface charges.}
The covariant phase space method yields the following expression for the surface charges associated with the diffeomorphisms $\xi$,
\begin{equation}
    k^{m n}[\xi]=\delta Q^{m n}[\xi]-2\Theta^{[m}\xi^{n]}\, .
\end{equation}
Varying $Q_{\xi}$ with respect to the metric tensor and using $\Theta$, one can find $k^{mn}$ as explicitly presented for GR and $f(R)$ as follows:
\begin{equation}
    k^{m n}_{_\text{GR}}[\xi]=\frac{\sqrt{-g}}{2}\left(h^{a[m}\nabla_{a}\xi^{n]}-\xi^{a}\nabla^{[m}h^{n]}_{a}-\frac{1}{2}h\nabla^{[m}\xi^{n]}+\xi^{[m}\nabla_{a}h^{n]a}-\xi^{[m}\nabla^{n]}h\right)\, ,
\end{equation}
and
\begin{small}
\begin{align}
k^{mn}_{\alpha_{0}} &= \Big(2h^{[m a}\nabla_{a}\xi^{n]}-2\nabla^{[m}h^{n] a}\xi_{a}-h\nabla^{[m}\xi^{n]}\Big)f^{\prime}+4\Big(R^{[m \alpha}\nabla_{\alpha}h-\nabla_{\alpha}R h^{[m\alpha}-R^{[m}_{\alpha}\nabla_{\beta}h^{\alpha\beta}\nonumber\\
&-\square\nabla^{[m}h+
\nabla_{\alpha}\nabla^{[m}\nabla_{\beta}h^{\alpha\beta}-
\nabla^{[m}\Big(R_{\alpha\beta}h^{\alpha\beta}\Big)+\dfrac{1}{2}\nabla^{[m}Rh\Big)\xi^{n]}f^{\prime\prime}+\delta R\nabla^{[m}\xi^{n]}f^{\prime\prime}-2\Theta_{f}^{[m}\xi^{n]}.
\end{align}
\end{small}
Finally, using \eqref{eqcharge},  one can obtain the variation of the conserved charge associated with a given Killing vector $\xi$. {Since the surface charge expressions are very lengthy, we do not include them here.}

\section{Quartic Gravity}
In this section, we apply the covariant phase space to the quartic terms. In an arbitrary dimension, there are 26 scalars constructed from algebraic combinations of four Riemann tensors \cite{Fulling:1992vm}. The action of a generic quartic theory is then given by 
\begin{align}\label{lagquart}
   I &= \dfrac{1}{2\kappa}\int d^{d}x\sqrt{-g}\Big\lbrace  \beta_{0}f(R)+\beta_{1}R^{2}R^{ab}R_{ab}+\beta_{2}RR^{ab}R_{a}^{c}R_{bc}+\beta_{3}R^{pq}R_{p}^{b}R_{q}^{s}R_{bs}\nonumber\\
   &+\beta_{4}RR^{pq}R^{bs}R_{pbqs} + \beta_{5}R^{pq}R^{bs}R_{b}^{a}R_{psqa}+\beta_{6}R^{2}R^{pqbs}R_{pqbs}+\beta_{7}R^{pq}R_{pq}R^{bsak}R_{bsak}\nonumber \\
   &+\beta_{8}R^{pq}R^{bs}R^{ak}{}_{pb}R_{akqs}+ \beta_{9}R^{pq}R^{bs}R^{a}{}_{p}{}^{k}{}_{q}R_{abks}+\beta_{10}R^{pq}R^{bsak}R_{bs}{}^{v}{}_{p}R_{akvq}+\beta_{11}\left(R^{pqbs}R_{pqbs}\right)^{2} \nonumber \\
   &+\beta_{12}R^{pqbs}R_{pq}{}^{ak}R_{ak}{}^{vw}R_{bsvw}+\beta_{13}R^{pqbs}R_{p}{}^{a}{}_{b}{}^{k}R_{a}{}^{v}{}_{k}{}^{w}R_{qvsw}+\beta_{14}\left(R_{ab}R^{ab}\right)^2\nonumber \\
   &+\beta_{15}RR^{pq}R^{bsa}{}_{p}R_{bsaq}+ \beta_{16}R^{pq}R_{p}^{b}R^{sak}{}_{q}R_{sakb}+\beta_{17}RR^{pqbs}R_{pq}{}^{ak}R_{bsak}+\beta_{18}RR^{pqbs}R_{p}{}^{a}{}_{b}{}^{k}R_{qask} \nonumber \\
   &+\beta_{19}R^{pqbs}R_{pqb}{}^{a}R^{kvw}{}_{s}R_{kvwa}+\beta_{20}R^{pqbs}R_{pq}{}^{ak}R_{ba}{}^{vw}R_{skvw}+\beta_{21}R^{pq}R^{bs}R^{a}{}_{p}{}^{k}{}_{b}R_{aqks}\nonumber \\
   &+\beta_{22}R^{pqbs}R_{kvsw}R_{p}{}^{a}{}_{b}{}^{k}R_{a}{}^{v}{}_{q}{}^{w}+\beta_{23}R^{pq}R_{p}{}^{b}{}_{q}{}^{s}R^{akv}{}_{b}R_{akvs}+\beta_{24}R^{pq}R^{bsak}R_{b}{}^{v}{}_{ap}R_{svkq}\nonumber \\
   &+\beta_{25}R^{pqbs}R_{pq}{}^{ak}R_{b}{}^{v}{}_{a}{}^{w}R_{svkw}
\Big\rbrace, 
\end{align}
where $\beta_{i}$ are the coupling constants. By variation of the Lagrangian \eqref{lagquart}, the equation of motion can be obtained, and we compactly provide the explicit expression in terms of the tensors $\mathcal{P}^{\alpha\beta\eta\delta}$ in Appendix \ref{fileq}.

\paragraph{Symplectic potential.}
Similarly, from the variation of the action for the quartic term, one can read the surface, $(d-1)$-form $\Theta$, which schematically leads to
\begin{equation}
\Theta^{m}=\sqrt{-g}\sum_{i=0}^{\text{25}}\Theta^{m}_{\beta_{i}}\ ,
\end{equation}
with the detailed expressions given in Appendix \ref{expressionsquartic}.
\paragraph{Noether-Wald charge.}
The Noether-Wald $(d-2)$-form $Q_{\xi}$ has the following structure 
\begin{equation}
    \mathcal{Q}^{mn}=\sqrt{-g}\sum_{i=1}^{\text{25}}Q^{mn}_{\beta_{i}}.
\end{equation}
The explicit expressions can be found in Appendix \ref{expressionsquartic}.
\paragraph{Surface charges.}
Consequently, the covariant phase-space method yields the following expression for the surface charges associated with the diffeomorphisms generated by, $\xi$,
\begin{equation}
    k^{mn}[\xi]=\delta Q^{mn}[\xi]-2\Theta^{[m}\xi^{n]}\, .
\end{equation}
Varying $Q_{\xi}$ with respect to the metric parameters and using $\Theta$, one can find $k^{mn}$.
Finally, using \eqref{eqcharge},  one can obtain the variation of the conserved charge associated with a given Killing vector $\xi$.

We have provided the explicit expressions for future reference, and now we move on to applications and analysis.

\section{Application in 3D Gravity}

To illustrate the general formalism developed above, we focus on a specific higher-curvature gravity model defined in \eqref{eqLAG}. This theory contains quadratic, cubic, and quartic curvature invariants and arises as a higher-order curvature deformation of New Massive Gravity (NMG), motivated by the holographic $c$-theorem within the AdS/CFT correspondence \cite{Sinha:2010ai}. Equivalently, the same action can be obtained from the infinitesimal curvature expansion of a Born–Infeld–like action, truncated at the corresponding order \cite{Afshar:2014ffa}. It has also been shown to exhibit universal features when the backreaction of a quantum conformally coupled scalar field is taken into account \cite{CieloQuantum}. The holographic structure and unitarity of the theory have been thoroughly investigated in \cite{Sajadi:2025wpx}. In this section, we compute the conserved charges and derive the first law(s) of thermodynamics for black hole solutions of the $(2+1)$-dimensional gravity theory \cite{Sinha:2010ai,Afshar:2014ffa,CieloHC}. The gravitational Lagrangian is given by
\begin{align}\label{eqLAG}
    \mathcal{L} &= R-2\Lambda+\eta\Big(R_{a b}R^{a b}-\dfrac{3}{8}R^2\Big)-\alpha\Big(\dfrac{17}{96}R^3-\dfrac{3}{4}RR_{ab}R^{ab}+\dfrac{2}{3}R_{a}{}^{b}R_{b}{}^{c}R_{c}{}^{a}\Big)+\beta_{0}R^4\nonumber\\
    &+\beta_{1}R^{2}R^{ab}R_{ab}+\beta_{2}RR^{ab}R_{a}^{c}R_{bc}+\beta_{3}R^{pq}R_{p}^{b}R_{q}^{s}R_{bs}+\beta_{14}\Big(R_{ab}R^{ab}\Big)^2.
\end{align}
The coupling constant $\beta_i$ can be written in terms of $\beta_0$ and $\beta$ as follows 
\begin{align}
    \beta_{1}=-\dfrac{17}{20}\beta-6\beta_{0},\;\;\;\beta_{2}=\dfrac{3}{5}\beta+8\beta_{0},\;\;\;\beta_{3}=-\dfrac{41}{20}\beta-6\beta_{0},\;\;\;\beta_{14}=\dfrac{21}{8}\beta+3\beta_{0},
\end{align}
where $\beta$ is a generic coupling constant of the quartic part of the action. As usual in higher curvature gravity, this theory admits more than one maximally symmetric solution for a generic value of the theory's parameters $\Lambda,\eta,\alpha$ and $\beta_i$. There exists generally four values of the effective cosmological constant for the solution depending on the parameters $\eta,\alpha,\beta$, and $\Lambda$, namely
\begin{equation}
   p(\Lambda_{\text{eff}})= \Lambda-\Lambda_{\text{eff}}-\dfrac{\eta\Lambda_{\text{eff}}^2}{4}-\dfrac{\alpha\Lambda_{\text{eff}}^3}{8}-\beta\Lambda_{\text{eff}}^4=0,
\end{equation}
leading to a quartic polynomial, where $\Lambda_{\text{eff}}$ is normalized such that on a maximally symmetric spacetime $R^{\mu\nu}_{\ \ \alpha\beta}=\Lambda_{\text{eff}}\left(\delta^\mu_\alpha\delta^\nu_\beta-\delta^\nu_\alpha\delta^\mu_\beta\right)$. We do not display the explicit roots, as they are lengthy and not very informative. It is interesting to notice that the constant $\beta_0$ does not appear in the determination of the effective cosmological constant and will neither appear in the expressions that follow. This is because, for the backgrounds under consideration, the component of field equations proportional to $\beta_0$ vanishes identically in (2+1) dimensions. 

Depending on the sign of the parameters, the solutions well-behave in the GR limit. 
For instance, when $\Lambda<0$, $\eta<0$, and $\beta>0$, three branches of the effective cosmological constant $\Lambda_{\text{eff}}$ smoothly approach the Einstein gravity value $\Lambda$ in the limit $\eta,\alpha,\beta \to 0$, while the remaining branch diverges.
There exists a special point in the parameter space at which the four maximally symmetric vacua become degenerate and coincide into a single solution ($p=p^{\prime}=p^{\prime\prime}=p^{\prime\prime\prime}=0$, where the primes denote the derivative with respect to $\Lambda_{eff}$). This point is given explicitly by
\begin{equation}
\Lambda_{\text{eff}}=-\frac{1}{2}\left(\frac{2}{\beta}\right)^{\frac{1}{3}},\qquad
\Lambda=-\frac{1}{8}\left(\frac{2}{\beta}\right)^{\frac{1}{3}},\qquad
\eta=6(4\beta)^{\frac{1}{3}},\qquad
\alpha=(32\beta)^{\frac{2}{3}} .
\end{equation}
For the following choice of parameters, the theory admits a unique AdS vacuum:
\begin{equation}\label{spepo}
  \Lambda=-\dfrac{1}{4\ell^2},\;\;\;\beta=\dfrac{\ell^6}{4},\;\;\;\eta=6\ell^2,\;\;\;\alpha=8\ell^4.
\end{equation}
and consequently leads to
\begin{equation}
\Lambda_{\text{eff}}=-\dfrac{1}{\ell^2} .
\end{equation}
We note that the relation between the couplings can be interpreted as natural. At the linearized level around the maximally symmetric vacuum, a new linearized gauge invariance emerges  (see e.g. \cite{Gabadadze:2012xv}).

\subsubsection{Near Horizon geometry, symmetries and charges}
{The covariant phase space method, which we applied in detail in the previous section to construct conserved charges associated with symmetries, {can likewise be used to systematically analyze near-horizon symmetries and derive their corresponding surface charge.} This framework was originally developed in the context of general relativity in \cite{NearHorizon1,NearHorizon2}, subsequently extended to New Massive Gravity \cite{NearHorizonNMG}, and more recently applied to higher dimensions within Lovelock gravity  \cite{NearHorizonEGB}.}

Adapting the Gaussian null coordinates $(v,\rho,\phi)$ to an event horizon in $(2+1)$ dimensions, the metric can be written as 
\begin{equation}
ds^2 = f\, dv^2 + 2k\, dv\, d\rho + 2h\, dv\, d\phi + R^2 d\phi^2,
\end{equation}
where the metric functions admit the near-horizon expansion,
\begin{align}\label{asympnh}
f &= -2\kappa \rho + \tau(\phi)\rho^2 + \mathcal{O}\!\left(\rho^3 \right), \nonumber\\
k &= 1 + \mathcal{O}\!\left(\rho^2 \right), \nonumber\\
h &= \theta(\phi)\rho + \sigma(\phi)\rho^2 + \mathcal{O}\!\left(\rho^3 \right), \nonumber\\
R^2 &= \Gamma^2(\phi) + \lambda(\phi)\rho + \mathcal{O}\!\left(\rho^2 \right).
\end{align}
{In these coordinates, $v$ is an advanced time coordinate that labels the null generators of the horizon, $\rho$ measures the affine distance away from the horizon and $\phi$ parametrizes the compact spatial cross section. Since $g_{\rho\rho}=0$, the hypersurfaces of constant $\rho$ are null and the event horizon is located at $\rho=0$.} The functions $\theta(\phi)$, $\Gamma(\phi)$, and $\lambda(\phi)$ are arbitrary functions of the angular coordinate $\phi$, which encodes possible angular deformations of the horizon data, while $\kappa$ identified with surface gravity, is considered constant. 

The near-horizon boundary conditions presented above are preserved by a class of asymptotic Killing vectors $\xi = \xi^{\mu}\partial_{\mu}$, which generate the symmetry algebra of the near-horizon phase space. These vectors implement supertranslations $P(\phi)$ and superrotations $L(\phi)$ along the horizon circle \cite{NearHorizon1}. Explicitly, they take the form
\begin{align}\label{horizonkilling}
\xi^{v} &= P(\phi) + \dots, \nonumber\\
\xi^{\rho} &= \frac{\theta(\phi)}{2\Gamma^2(\phi)}\,\partial_{\phi}P(\phi)\,\rho^2 + \dots, \nonumber\\
\xi^{\phi} &= L(\phi) - \frac{1}{\Gamma^{2}(\phi)}\,\partial_{\phi}P(\phi)\,\rho 
+ \frac{\lambda(\phi)}{2\Gamma^{4}(\phi)}\,\partial_{\phi}P(\phi)\,\rho^2 + \dots,
\end{align}
where the ellipses denote the subleading terms in the expansion around $\rho \to 0$. These vector fields preserve the form of the near-horizon metric and define the asymptotic symmetry generators associated with the horizon degrees of freedom.

\paragraph{Symplectic potential.} We begin with the symplectic potential. Upon restricting to the solution space and imposing the near-horizon boundary conditions, we find that the radial component vanishes,
\begin{equation}\label{symp-pot-Eins-formal}
   \Theta^{\rho}[\delta g;g]=0 .
\end{equation}
This follows from the fact that, on the horizon $\rho=0$, the allowed variations preserve the form of the metric and are independent of $v$, so that all leading contributions to $\Theta^{\rho}$ vanish under the imposed fall-off conditions. Since the symplectic potential vanishes on the horizon under the proposed boundary conditions, the central term in the charge algebra is absent, and the asymptotic symmetry algebra admits no central extension.

\paragraph{Noether-Wald charge.}

The Noether--Wald charge associated with the vector field $\xi$ can be written in the form
\begin{align}\label{NW-charge-compact}
    Q^{v\rho}[\xi]
    =
    P(\phi)\,Q^{P}
    +
    L(\phi)\,Q^{L},
\end{align}
where the charges admit the following simple structure:
\begin{align}
    Q^{P}
    &=
    \kappa \Bigg[
        2\Gamma
        + \frac{\eta}{\Gamma^2}\,\mathcal{A}_0
        + \frac{\alpha}{2\Gamma^4}\,\mathcal{X}\,\mathcal{A}_1
        + \frac{24\beta}{5\Gamma^6}\,\mathcal{X}^2\,\mathcal{A}_2
    \Bigg],
\\[6pt]
    Q^{L}
&=
-\!\left[\Gamma
\theta
+\frac{\eta}{8\Gamma^{2}}\,\mathcal{B}_0
+\frac{2\alpha}{\Gamma^{4}}\,\mathcal{X}\,\mathcal{B}_1
+\frac{12\beta}{5\Gamma^{6}}\,\mathcal{X}^{2}\,\mathcal{B}_2
\right].
\end{align}
where we have defined the repeated geometric structure
\begin{equation}
    \mathcal{X}(\phi)
    :=
    \tau(\phi)\Gamma^2(\phi)
    -
    \frac{1}{4}\theta^2(\phi),
\end{equation}
and the coefficient functions
\begin{align}
    \mathcal{A}_0
    &=
    \theta\,\partial_{\phi}\Gamma
    + \Gamma\!\left(
         \kappa\lambda
        - \partial_{\phi}\theta
        + \frac{1}{4}\theta^2+\tau\Gamma^2
    \right),
\\
    \mathcal{A}_1
    &=
    \theta\,\partial_{\phi}\Gamma
    + \Gamma\!\left(
        \kappa\lambda
        - \partial_{\phi}\theta
        + \frac{1}{8}\theta^2
        + \frac{3}{2}\tau\Gamma^2
    \right),
\\
    \mathcal{A}_2
    &=
    \theta\,\partial_{\phi}\Gamma
    + \Gamma\!\left(
        \kappa\lambda
        - \partial_{\phi}\theta
        + \frac{1}{12}\theta^2
        + \frac{5}{3}\tau\Gamma^2
    \right),
\end{align}
and
\begin{align}
   \mathcal{B}_0
&=
4\,\theta^2\,\partial_{\phi}\Gamma
+\Gamma\!\left(
    -4\theta\,\partial_{\phi}\theta
    +\theta^3
    +4\kappa\lambda\theta
    +\Gamma^2\!\big(
        -16\,\partial_{\phi}\tau
        +20\,\tau\theta
        +32\,\kappa\sigma
    \big)
\right),
\\
    \mathcal{B}_1
    &=
    \frac{1}{8}\theta^2\partial_{\phi}\Gamma
    + \Gamma\!\left(
        -\frac{1}{8}\theta\partial_{\phi}\theta
        -\frac{1}{2}\Gamma^2\partial_{\phi}\tau
        +\Gamma^2\!\left(
            \kappa\sigma
            + \frac{11}{16}\tau\theta
        \right)
        +\frac{1}{8}\kappa\lambda\theta
        +\frac{1}{64}\theta^3
    \right),
\\
    \mathcal{B}_2
    &=
    \theta^2\partial_{\phi}\Gamma
    + \Gamma\!\left(
        -\theta\partial_{\phi}\theta
        -4\Gamma^2\partial_{\phi}\tau
        +\Gamma^2\!\left(
            8\kappa\sigma
            + \frac{17}{3}\tau\theta
        \right)
        +\kappa\lambda\theta
        +\frac{1}{12}\theta^3
    \right).
\end{align}
{Notice that $Q^{v\rho}$ has to be integrated on the horizon in order to obtain the actual charge.}

It is convenient to express the charges in terms of the corresponding Einstein charges \cite{NearHorizon1} as follows:
\begin{align}
Q^{P}
&=
Q^{P}_{\text{Einstein}}
\left[
1
+\frac{\eta}{2\Gamma^{3}}\,\mathcal{A}_0
+\frac{\alpha}{4\Gamma^{5}}\,\mathcal{X}\,\mathcal{A}_1
+\frac{12\beta}{5\Gamma^{7}}\,\mathcal{X}^{2}\,\mathcal{A}_2
\right],
\\[6pt]
Q^{L}
&=
Q^{L}_{\text{Einstein}}
\left[
1
+\frac{\eta}{8\Gamma^{3}\theta}\,\mathcal{B}_0
+\frac{2\alpha}{\Gamma^{5}\theta}\,\mathcal{X}\,\mathcal{B}_1
+\frac{12\beta}{5\Gamma^{7}\theta}\,\mathcal{X}^{2}\,\mathcal{B}_2
\right],
\end{align}
where
\[
Q^{P}_{\text{Einstein}}=2\kappa \Gamma,
\qquad
Q^{L}_{\text{Einstein}}=-\Gamma\theta.
\]
Both charges can be unified into the single expression
\begin{align}
Q^{I}
=
Q^{I}_{\text{Einstein}}
\left[
1
+\sum_{n=0}^{2}
C^{I}_{n}\,
\Gamma^{-(2n+3)}\,
\Theta_I^{-1}\,
\mathcal{X}^{\,n}\,
\mathcal{F}^{I}_{n}
\right],
\qquad I\in\{P,L\},
\end{align}
with
\[
\Theta_P = 1,
\qquad
\mathcal{F}^{P}_{n} = \mathcal{A}_n,
\qquad
(C^{P}_{0},C^{P}_{1},C^{P}_{2})
=
\left(
\tfrac{\eta}{2},\,
\tfrac{\alpha}{4},\,
\tfrac{12\beta}{5}
\right),
\]
\[
\Theta_L = \theta,
\qquad
\mathcal{F}^{L}_{n} = \mathcal{B}_n,
\qquad
(C^{L}_{0},C^{L}_{1},C^{L}_{2})
=
\left(
\tfrac{\eta}{8},\,
2\alpha,\,
\tfrac{12\beta}{5}
\right).
\]

We can also rewrite the above expressions entirely in terms of the near-horizon data as follows:
\begin{align}
Q^{P}
&=
2\kappa\left[
\Gamma
+\sum_{n=0}^{2}
\frac{c_{n}\mathcal{X}^{n}}{2\Gamma^{2(n+1)}}
\left(
y
+\frac{2n+1}{n+1}\,\tau \Gamma^{3}
+\frac{1}{4(n+1)}\,\theta^{2}\Gamma
\right)
\right],
\\
Q^{L}
&=
-\left[
\theta\Gamma
+\sum_{n=0}^{2}
\frac{c_{n}\mathcal{X}^{n}}{2\Gamma^{2(n+1)}}
\left(
\hat{y}
+\frac{2n+1}{n+1}\,\tau \theta\Gamma^{3}
+\frac{1}{4(n+1)}\,\theta^{3}\Gamma
\right)
\right],
\end{align}
where
\begin{equation}
c_{0}=\eta,\;\;
c_{1}=\frac{\alpha}{2},\;\;
c_{2}=\frac{24\beta}{5},\quad
y=\theta\,\partial_{\phi}\Gamma+\Gamma\!\left(\kappa\lambda-\partial_{\phi}\theta\right),\quad
\hat{y}= \theta y+{4\Gamma^{3}}\left(2\kappa\sigma-\partial_{\phi}\tau\right).
\end{equation}
{These expressions suggest a natural generalization of the near-horizon charges for the family of theories considered in this paper, to arbitrarily higher powers of the Riemann tensor.}

Let us now consider the case where all near-horizon fields are constant along the angular direction $\phi$, namely
\[
\partial_{\phi}\Gamma
=
\partial_{\phi}\theta
=
\partial_{\phi}\tau
=
\partial_{\phi}\lambda
=
\partial_{\phi}\sigma
=
0 .
\]
{This restricted case will be sufficiently general as to encompass the static and rotating backgrounds considered below.}

In this case, all derivative contributions vanish, and the structure simplifies considerably. The charges therefore reduce to their zero modes:
\begin{align}
Q^{P}_{0}
&=
2\kappa
\left[
\Gamma
+
\sum_{n=0}^{2}
\frac{c_n \mathcal{X}^n}{2\Gamma^{2(n+1)}}
\left(
\kappa\Gamma\lambda
+
\frac{2n+1}{n+1}\tau\Gamma^3
+
\frac{1}{4(n+1)}\theta^2\Gamma
\right)
\right],
\\[6pt]
Q^{L}_{0}
&=
-
\left[
\theta\Gamma
+
\sum_{n=0}^{2}
\frac{c_n \mathcal{X}^n}{2\Gamma^{2(n+1)}}
\left(
\theta\kappa\Gamma\lambda
+
{8\kappa\sigma\Gamma^3}
+
\frac{2n+1}{n+1}\theta\tau\Gamma^3
+
\frac{1}{4(n+1)}\theta^3\Gamma
\right)
\right].
\end{align}
{In order to construct the final charge, out of these constant quantities, we must multiply them by $2\pi$ from the integration along the periodic $\phi$ coordinate.} {Thus, in the constant-field sector, the charges reduce to global quantities determined by the near-horizon data. 
The charges \(Q^{P}\) and \(Q^{L}\) describe the symmetries near the black hole horizon and the associated soft hair. These charges characterize the horizon degrees of freedom and can be used to distinguish different microscopic states of the black hole, suggesting that the entropy originates from the large number of such states. Accordingly, the entropy is proportional to \(Q^{P}/\kappa\), which reproduces the Bekenstein--Hawking result in Einstein gravity and includes additional corrections in the presence of higher-curvature terms \cite{Hawking:2016msc,Donnay:2015abr,Afshar:2016wfy,Afshar:2017okz}.
The charge \(Q^{L}\) measures the rotation of the black hole horizon and therefore corresponds to the black hole’s angular momentum ($Q^{L}_{0}=\vert J\vert$). 
Accordingly, we can rewrite the first law and the Smarr relation in terms of the near-horizon charges as}
\begin{equation}
\delta M
=
\frac{\kappa}{2\pi}\,
\delta\!\left(\frac{Q^{P}_{0}}{\kappa}\right)
-
\Omega_H\,\delta Q^{L}_{0},
\qquad
M \propto Q^{P}_{0} - \Omega_H Q^{L}_{0}.
\end{equation}

{As a final remark, we note that the near-horizon geometry obtained here naturally carries a Carrollian structure, reflecting the degeneracy of the induced metric on the non-expanding null horizon \cite{Donnay:2019jiz}. In this framework, the charges $Q_P$ and $Q_L$ can be interpreted as the energy and momentum densities of a Carrollian fluid, and their zero modes reproduce the thermodynamic quantities $TS$ and $J$, respectively. Thus, the near-horizon symmetries and conserved charges derived in this work capture the effective horizon dynamics of higher-curvature gravity.}

In the following, we apply this formalism to the near horizon of the rotating BTZ black hole, 
as well as to static and rotating hairy black holes \cite{Banados:1992wn,Donnay:2020yxw,OTT}.

\paragraph{Rotating BTZ Black Hole.}

Let us now evaluate the charges for the near-horizon of the rotating BTZ  \cite{Donnay:2020yxw}
\begin{equation}\label{eqq39}
\kappa=\frac{r_{+}^2-r_{-}^2}{\ell^2 r_{+}},
\quad
\Gamma=r_{+},
\quad
\theta=\frac{2r_{-}}{\ell},
\quad
\lambda=2 r_{+},
\quad
\tau=-\frac{r_{+}^2-r_{-}^2}{\ell^2 r_{+}^2},
\quad
\sigma=\frac{r_{-}}{\ell r_{+}} .
\end{equation}
The BTZ BH is a solution to the field equations of the above theory, provided that
\begin{equation}\label{eqsp}
\Lambda=-\dfrac{1}{\ell^2}+\dfrac{\eta}{4\ell^4}+\dfrac{\beta}{\ell^8}-\dfrac{\alpha}{8\ell^6}.
\end{equation}
The charges can then be written as
\begin{align}
Q^{P}_{0}
=&
\frac{2\left(r_{+}^{2}-r_{-}^{2}\right)}{\ell^{2}}
\left(
1+\frac{\eta}{2\ell^{2}}
-\frac{\alpha}{8\ell^{4}}
+\frac{4\beta}{5\ell^{6}}
\right),\\
Q^{L}_{0}
=&
-\frac{2 r_{+} r_{-}}{\ell}
\left(
1+\frac{\eta}{2\ell^{2}}
-\frac{\alpha}{8\ell^{4}}
+\frac{4\beta}{5\ell^{6}}
\right).
\end{align}
Therefore, both charges are proportional to their Einstein gravity
counterparts,
\begin{equation}
Q^{P,L}_{0}
=
Q^{P,L}_{\text{Einstein}}
\left(
1+\frac{\eta}{2\ell^{2}}
-\frac{\alpha}{8\ell^{4}}
+\frac{4\beta}{5\ell^{6}}
\right).
\end{equation}
It is easy to show that $Q^{P}_{0}=TS$ and $Q_{0}^{L}=J$. The above formulas correctly reproduce black hole entropy $S$ and angular momentum $J$ from the near-horizon perspective, in the higher curvature gravity up to quartic order, which we considered in this section.

\paragraph{Static Hairy Black Hole.}

For the near-horizon geometry of the hairy black hole, one finds \cite{OTT} (see also \cite{BHTMore,OTTQNM,OTTLeston})
\begin{align}\label{eqq60}
\kappa &= \dfrac{r_{+}-r_{-}}{2\ell^2}, \hspace{0.5cm}
\tau = -\dfrac{1}{\ell^2}, \hspace{0.5cm}
\Gamma = r_{+}, \hspace{0.5cm}
\lambda = 2r_{+}, \hspace{0.5cm}
\theta = 0 .
\end{align}
The hairy black hole is a solution of the field equations of the above theory, provided that
\begin{equation}\label{eqsp1}
\Lambda = -\dfrac{1}{\ell^2}
+ \dfrac{\beta}{\ell^8}
+ \dfrac{\eta}{4\ell^4}
- \dfrac{\alpha}{8\ell^6},
\qquad
\eta = 2\ell^2
+ \dfrac{3\alpha}{4\ell^2}
- \dfrac{8\beta}{\ell^4}.
\end{equation}
For the background with $\theta=0$, the Einstein contribution to $Q^{L}$ vanishes identically, and all higher-curvature corrections reduce to proportional factors of this term. Consequently, as expected, we obtain $Q^{L}=0$, implying that the solution carries no angular momentum, $J=0$, and therefore corresponds to a non-rotating configuration.
The charge associated with the zero modes of $P$ leads to
\[
Q^{P}
=
\frac{(r_{+}-r_{-})^{2}}{8\ell^{2}}
\left(
1+\frac{\alpha}{8\ell^{4}}
-\frac{8\beta}{5\ell^{6}}
\right).
\]

Using the near-horizon expansion, the above formula reproduces the entropy of a black hole ($Q^{P}=TS$).
Unlike the BTZ solution, the higher-curvature terms do not simply renormalize the Einstein contribution $Q^{P}\neq Q^{P}_{\text{Einstein}}\left(
1+\frac{\alpha}{8\ell^{4}}
-\frac{8\beta}{5\ell^{6}}
\right)$, {which is due to the presence of the hair of gravitational origin.}

\paragraph{Rotating Hairy Black Hole.}

The near-horizon data of the rotating generalization of the hairy black hole can be expressed in terms of the outer and inner horizon radii $r_+$ and $r_-$ as
\begin{align}
\kappa &=
\frac{\sqrt{\zeta}}{\ell^2\sqrt{2(1+\zeta)}}(r_+-r_-),
\qquad
\theta =
\frac{r_+-r_-}{\ell}
\sqrt{\frac{1-\zeta}{2\zeta}},
\qquad
\Gamma =
\frac{\ell}{2}(r_+-r_-), \nonumber\\[6pt]
\lambda &=
r_+\sqrt{\frac{1+\zeta}{2\zeta}},
\qquad
\tau = \frac{1}{\ell^2},
\qquad
\sigma =
\frac{a}{2\ell^2}
\frac{(r_+-r_-)}{(r_++r_-)^2}.
\end{align}
Here $\zeta=\sqrt{1-a^2/\ell^2}$, with $a$ denoting the rotation parameter.
The corresponding conserved charges read
\begin{equation}
Q^{P}=
\frac{(r_{+}-r_{-})^{2}}{8\ell^{2}}
\left(
1+\frac{\alpha}{8\ell^{4}}
-\frac{8\beta}{5\ell^{6}}
\right),
\qquad 
Q^{L} =\ell\sqrt{\frac{1-\zeta}{\zeta}}\,Q^{P}.
\end{equation}
The charge $Q^{P}$ scales with the square of the horizon separation and receives higher-curvature corrections through the universal multiplicative factor in parentheses. 
The second charge $Q^{L}$ is proportional to $Q^{P}$, with the rotation-dependent factor 
$\ell\sqrt{\frac{1-\zeta}{\zeta}}$ ($\zeta=\sqrt{1-a^{2}/\ell^{2}}$), 
so that $Q^{L}\to0$ in the static limit and $Q^{L}\simeq a\,Q^{P}$ for small rotation.

\section{Conclusion}\label{sec3}

In this work, we first review the solution phase space method and then apply it to a broad class of higher-curvature gravitational theories. These include quartic, cubic, and quadratic gravities, as well as theories with the Ricci scalar terms. After presenting the symplectic potential and the Noether-Wald charge, the key ingredients for the computation, we analyze two families of black hole solutions, i.e., BTZ and hairy black holes. In particular, we evaluate their conserved charges and verify their consistency with results previously obtained by alternative approaches. The main strengths of this method are its universality across arbitrary higher-curvature theories and dimensions also its independence from asymptotic structure or horizon properties, and the straightforward proof of the first law of black hole thermodynamics.

\section*{Acknowledgements} 
This research has received funding support from the NSRF via the Program Management Unit for Human Resource and Institutional Development, Research and Innovation grant number $B13F680083$. The work of J.O. is partially funded by FONDECYT GRANT 1262452.

\appendix

\section{Field Equations for Cubic and Quartic Part}\label{fileq}
By variation of the action with respect to the metric tensor, one can obtain the corresponding equation of motion as follows
\begin{align}
\mathcal{E}_{ab}=\dfrac{1}{\sqrt{-g}}\dfrac{\delta(\sqrt{-g}\mathcal{L})}{\delta g^{ab}}=\mathcal{P}_{acde}\mathcal{R}_{b}{}^{cde}-\dfrac{1}{2}g_{ab}\mathcal{L}-2\nabla^{c}\nabla^{d}\mathcal{P}_{acdb}, \label{LHS}
\end{align}
here $\mathcal{P}^{abcd}=\dfrac{\partial\mathcal{L}}{\partial\mathcal{R}_{abcd}}$. 
For the cubic gravity, we have
\begin{align}
\mathcal{P}^{\alpha\beta\eta\delta}_{\alpha_{0}}&=\dfrac{1}{2}\left(g^{\alpha\eta}g^{\beta\delta}-g^{\alpha\delta}g^{\beta\eta}\right)f^{\prime}(R),\\
    \mathcal{P}^{\alpha\beta\eta\delta}_{\alpha_{1}}&=-\dfrac{3}{4}\Big(g^{\beta\eta}R^{\alpha\mu}R^{\delta}_{\mu}-g^{\alpha\eta}R^{\beta\mu}R^{\delta}_{\mu}-g^{\beta\delta}R^{\alpha\mu}R^{\eta}_{\mu}+g^{\alpha\delta}R^{\beta\mu}R^{\eta}_{\mu}\Big),\\
    \mathcal{P}^{\alpha\beta\eta\delta}_{\alpha_{2}}&=\dfrac{1}{2}g^{\alpha\eta}g^{\beta\delta}R_{\mu\nu}R^{\mu\nu}-\dfrac{1}{2}g^{\alpha\delta}g^{\beta\eta}R_{\mu\nu}R^{\mu\nu}-\dfrac{1}{2}g^{\beta\eta}g^{\alpha\delta}R+\dfrac{1}{2}g^{\beta\delta}R^{\alpha\eta}R+\dfrac{1}{2}g^{\alpha\eta}R^{\beta\delta}R\nonumber\\
    &-\dfrac{1}{2}g^{\alpha\delta}R^{\beta\eta}R,\\
    \mathcal{P}^{\alpha\beta\eta\delta}_{\alpha_{3}}&=-3R^{\alpha\beta\mu\nu}R^{\delta\eta}{}_{\mu\nu},\\
    \mathcal{P}^{\alpha\beta\eta\delta}_{\alpha_{4}}&=\dfrac{1}{2}R^{\alpha\eta}R^{\beta\delta}-\dfrac{1}{2}R^{\alpha\delta}R^{\beta\eta}-\dfrac{1}{2}g^{\beta\eta}R^{\mu\nu}R^{\alpha}{}_{\mu}{}^{\delta}{}_{\nu}+\dfrac{1}{2}g^{\beta\delta}R^{\mu\nu}R^{\alpha}{}_{\mu}{}^{\eta}{}_{\nu}+\dfrac{1}{2}g^{\alpha\eta}R^{\mu\nu}R^{\beta}{}_{\mu}{}^{\delta}{}_{\nu}\nonumber\\
    &-\dfrac{1}{2}g^{\alpha\delta}R^{\mu\nu}R^{\beta}{}_{\mu}{}^{\eta}{}_{\nu},\\
    \mathcal{P}^{\alpha\beta\eta\delta}_{\alpha_{5}}&=\dfrac{1}{2}g^{\alpha\eta}g^{\beta\delta}R_{\mu\nu\rho\sigma}R^{\mu\nu\rho\sigma}-\dfrac{1}{2}g^{\alpha\delta}g^{\beta\eta}R_{\mu\nu\rho\sigma}R^{\mu\nu\rho\sigma}-2 R R^{\alpha\beta\delta\eta},\\
    \mathcal{P}^{\alpha\beta\eta\delta}_{\alpha_{6}}&=\dfrac{1}{2}R^{\beta\mu}R^{\alpha}{}_{\mu}{}^{\delta\eta}+\dfrac{1}{2}R^{\eta\mu}R^{\alpha\beta\delta}{}_{\mu}-\dfrac{1}{2}R^{\delta\mu}R^{\alpha\beta\eta}{}_{\mu}-\dfrac{1}{2}R^{\alpha\mu}R^{\beta}{}_{\mu}{}^{\delta\eta}+\dfrac{1}{4}g^{\beta\eta}R^{\alpha\mu\nu\rho}R^{\delta}{}_{\mu\nu\rho}\nonumber\\
    &-\dfrac{1}{4}g^{\alpha\eta}R^{\beta\mu\nu\rho}R^{\delta}{}_{\mu\nu\rho}-\dfrac{1}{4}g^{\beta\delta}R^{\alpha\mu\nu\rho}R^{\eta}{}_{\mu\nu\rho}+\dfrac{1}{4}g^{\alpha\delta}R^{\beta\mu\nu\rho}R^{\eta}{}_{\mu\nu\rho},\\
    \mathcal{P}^{\alpha\beta\eta\delta}_{\alpha_{7}}&=\dfrac{3}{2}R^{\alpha\mu\eta\nu}R^{\beta}{}_{\mu}{}^{\delta}{}_{\nu}-\dfrac{3}{2}R^{\alpha\mu\delta\nu}R^{\beta}{}_{\mu}{}^{\eta}{}_{\nu}.
\end{align}
For quartic gravity, we have
\begin{align}
 \mathcal{P}^{\alpha\beta\eta\delta}_{\beta_{1}}&=g^{\alpha\eta}g^{\beta\delta}R_{\mu\nu}R^{\mu\nu}R-g^{\alpha\delta}g^{\beta\eta}R_{\mu\nu}R^{\mu\nu}R-\dfrac{1}{2}g^{\beta\eta}R^{\alpha\delta}R^{2}+\dfrac{1}{2}g^{\beta\delta}R^{\alpha\eta}R^{2}+\dfrac{1}{2}g^{\alpha\eta}R^{\beta\delta}R^2\nonumber\\
 &-\dfrac{1}{2}g^{\alpha\delta}R^{\beta\eta}R^{2},\\
 \mathcal{P}^{\alpha\beta\eta\delta}_{\beta_{2}}&=\dfrac{1}{2}g^{\alpha\eta}g^{\beta\delta}R_{\mu}^{\rho}R^{\mu\nu}R_{\nu\rho}-\dfrac{1}{2}g^{\alpha\delta}g^{\beta\eta}R_{\mu}^{\rho}R^{\mu\nu}R_{\nu\rho}-\dfrac{3}{4}g^{\beta\eta}R^{\alpha\mu}R^{\delta}_{\mu}R+\dfrac{3}{4}g^{\alpha\eta}R^{\beta\mu}R^{\delta}_{\mu}R\nonumber\\
 &+\dfrac{3}{4}g^{\beta\delta}R^{\alpha\mu}R^{\eta}_{\mu}R-\dfrac{3}{4}g^{\alpha\delta}R^{\beta\mu}R^{\eta}_{\mu}R, \\
\mathcal{P}^{\alpha\beta\eta\delta}_{\beta_{3}}&=-g^{\beta\eta}R_{\mu\nu}R^{\alpha\mu}R^{\delta\nu}+g^{\alpha\eta}R_{\mu\nu}R^{\beta\mu}R^{\delta\nu}+g^{\beta\delta}R_{\mu\nu}R^{\alpha\mu}R^{\eta\nu}-g^{\alpha\delta}R_{\mu\nu}R^{\beta\mu}R^{\eta\nu}, \\
\mathcal{P}^{\alpha\beta\eta\delta}_{\beta_{4}}&=\dfrac{1}{2}R^{\alpha\eta}R^{\beta\delta}R-\dfrac{1}{2}R^{\alpha\delta}R^{\beta\eta}R+\dfrac{1}{2}g^{\alpha\eta}g^{\beta\delta}R^{\mu\nu}R^{\rho\sigma}R_{\mu\rho\nu\sigma}-\dfrac{1}{2}g^{\alpha\delta}g^{\beta\eta}R^{\mu\nu}R^{\rho\sigma}R_{\mu\rho\nu\sigma}\nonumber\\
     &-\dfrac{1}{2}g^{\beta\eta}R^{\mu\nu}RR^{\alpha}{}_{\mu}{}^{\delta}{}_{\nu}+\dfrac{1}{2}g^{\beta\delta}R^{\mu\nu}R R^{\alpha}{}_{\mu}{}^{\eta}{}_{\nu}+\dfrac{1}{2}g^{\alpha\eta}R^{\mu\nu}R R^{\beta}{}_{\mu}{}^{\delta}{}_{\nu}-\dfrac{1}{2}g^{\alpha\delta}R^{\mu\nu}R R^{\beta}{}_{\mu}{}^{\eta}{}_{\nu},\\ 
\mathcal{P}^{\alpha\beta\eta\delta}_{\beta_{5}}&=\dfrac{1}{4}R^{\alpha\eta}R^{\beta\mu}R_{\mu}^{\delta}-\dfrac{1}{4}R^{\alpha\mu}R^{\beta\eta}R_{\mu}^{\delta}-\dfrac{1}{4}R^{\alpha\delta}R^{\beta\mu}R_{\mu}^{\eta}+\dfrac{1}{4}R^{\alpha\mu}R^{\beta\delta}R^{\eta}_{\mu}-\dfrac{1}{4}g^{\beta\eta}R^{\nu\rho}R^{\delta\mu}R^{\alpha}{}_{\nu\mu\rho}\nonumber\\
     &+\dfrac{1}{4}g^{\beta\delta}R^{\nu\rho}R^{\eta\mu}{R^{\alpha}}_{\nu\mu\rho}-\dfrac{1}{4}g^{\beta\eta}R_{\mu}^{\rho}R^{\mu\nu}R^{\alpha}{}_{\nu}{}^{\delta}{}_{\rho}+\dfrac{1}{4}g^{\beta\delta}R_{\mu}^{\rho}R^{\mu\nu}R^{\alpha}{}_{\nu}{}^{\eta}{}_{\rho}+\dfrac{1}{4}g^{\alpha\eta}R^{\nu\rho}R^{\delta\mu}R^{\beta}{}_{\nu\mu\rho}\nonumber\\
     &-\dfrac{1}{4}g^{\alpha\delta}R^{\nu\rho}R^{\eta\mu}R^{\beta}{}_{\nu\mu\rho}+\dfrac{1}{4}g^{\alpha\eta}R_{\mu}^{\rho}R^{\mu\nu}R^{\beta}{}_{\nu}{}^{\delta}{}_{\rho}-\dfrac{1}{4}g^{\alpha\delta}R_{\mu}^{\rho}R^{\mu\nu}R^{\beta}{}_{\nu}{}^{\eta}{}_{\rho}-\dfrac{1}{4}g^{\beta\eta}R^{\nu\rho}R^{\alpha\mu}R^{\delta}{}_{\nu\mu\rho}\nonumber\\
     &+\dfrac{1}{4}g^{\alpha\eta}R^{\nu\rho}R^{\beta\mu}R^{\delta}{}_{\nu\mu\rho}+\dfrac{1}{4}g^{\beta\delta}R^{\nu\rho}R^{\alpha\mu}R^{\eta}{}_{\nu\mu\rho}-\dfrac{1}{4}g^{\alpha\delta}R^{\nu\rho}R^{\beta\mu}R^{\eta}{}_{\nu\mu\rho},\\
     \mathcal{P}^{\alpha\beta\eta\delta}_{\beta_{6}}&=g^{\alpha\eta}g^{\beta\delta}RR_{\mu\nu\rho\sigma}R^{\mu\nu\rho\sigma}-g^{\alpha\delta}g^{\beta\eta}RR_{\mu\nu\rho\sigma}R^{\mu\nu\rho\sigma}-2R^2R^{\alpha\beta\delta\eta},\\
     \mathcal{P}^{\alpha\beta\eta\delta}_{\beta_{7}}&=-\dfrac{1}{2}g^{\beta\eta}R^{\alpha\delta}R_{\mu\nu\rho\sigma}R^{\mu\nu\rho\sigma}+\dfrac{1}{2}g^{\beta\delta}R^{\alpha\eta}R_{\mu\nu\rho\sigma}R^{\mu\nu\rho\sigma}+\dfrac{1}{2}g^{\alpha\eta}g^{\beta\delta}R_{\mu\nu\rho\sigma}R^{\mu\nu\rho\sigma}\nonumber\\
     &-\dfrac{1}{2}g^{\alpha\delta}R^{\beta\eta}R_{\mu\nu\rho\sigma}R^{\mu\nu\rho\sigma}-2R_{\mu\nu}R^{\mu\nu}R^{\alpha\beta\delta\eta},\\
     \mathcal{P}^{\alpha\beta\eta\delta}_{\beta_{8}}&=-R^{\delta\mu}R^{\eta\nu}R^{\alpha\beta}{}_{\mu\nu}-\dfrac{1}{2}g^{\beta\eta}R^{\mu\nu}R^{\alpha}{}_{\mu}{}^{\rho\sigma}R^{\delta}{}_{\nu\rho\sigma}+\dfrac{1}{2}g^{\alpha\eta}R^{\mu\nu}R^{\beta}{}_{\mu}{}^{\rho\sigma}R^{\delta}{}_{\nu\rho\sigma}-R^{\alpha\mu}R^{\beta\nu}R^{\delta\eta}{}_{\mu\nu}\nonumber\\
     &+\dfrac{1}{2}g^{\beta\delta}R^{\mu\nu}R^{\alpha}{}_{\mu}{}^{\rho\sigma}R^{\eta}{}_{\nu\rho\sigma}-\dfrac{1}{2}g^{\alpha\delta}R^{\mu\nu}R^{\beta}{}_{\mu}{}^{\rho\sigma}R^{\eta}{}_{\nu\rho\sigma},\\
     \mathcal{P}^{\alpha\beta\eta\delta}_{\beta_{9}}&=-\dfrac{1}{2}R^{\mu\nu}R^{\beta\eta}R^{\alpha}{}_{\mu}{}^{\delta}{}_{\nu}+\dfrac{1}{2}R^{\mu\nu}R^{\beta\delta}R^{\alpha}{}_{\mu}{}^{\eta}{}_{\nu}-\dfrac{1}{2}g^{\beta\eta}R^{\mu\nu}R_{\mu\rho\nu\sigma}R^{\alpha\rho\delta\sigma}+\dfrac{1}{2}R^{\mu\nu}R^{\alpha\eta}R^{\beta}{}_{\mu}{}^{\delta}{}_{\nu}\nonumber\\
     &+\dfrac{1}{2}g^{\beta\delta}R^{\mu\nu}R_{\mu\rho\nu\sigma}R^{\alpha\rho\eta\sigma}-\dfrac{1}{2}R^{\mu\nu}R^{\alpha\delta}R^{\beta}{}_{\mu}{}^{\eta}{}_{\nu}+\dfrac{1}{2}g^{\alpha\eta}R^{\mu\nu}R_{\mu\rho\nu\sigma}R^{\beta\rho\delta\sigma}-\dfrac{1}{2}g^{\alpha\delta}R^{\mu\nu}R_{\mu\rho\nu\sigma}R^{\beta\rho\eta\sigma}, \\          
\mathcal{P}^{\alpha\beta\eta\delta}_{\beta_{10}}&=-\dfrac{1}{2}R^{\eta\mu}R^{\alpha\beta\nu\rho}R^{\delta}{}_{\mu\nu\rho}-\dfrac{1}{4}g^{\beta\eta}R_{\nu\rho\sigma\lambda}R^{\alpha\mu\nu\rho}R^{\delta}{}_{\mu}{}^{\sigma\lambda}+\dfrac{1}{4}g^{\alpha\eta}R_{\nu\rho\sigma\lambda}R^{\beta\mu\nu\rho}R^{\delta}{}_{\mu}{}^{\sigma\lambda}\nonumber\\
    &-R^{\mu\nu}R^{\alpha\beta}{}_{\mu}{}^{\rho}R^{\delta\eta}{}_{\nu\rho}-\dfrac{1}{2}R^{\beta\mu}R^{\alpha}{}_{\mu\nu\rho}R^{\delta\eta\nu\rho}+\dfrac{1}{2}R^{\alpha\mu}R^{\beta}{}_{\mu\nu\rho}R^{\delta\eta\nu\rho}+\dfrac{1}{2}R^{\delta\mu}R^{\alpha\beta\nu\rho}R^{\eta}{}_{\mu\nu\rho}\nonumber\\
    &+\dfrac{1}{4}g^{\beta\delta}R_{\nu\rho\sigma\lambda}R^{\alpha\mu\nu\rho}R^{\eta}{}_{\mu}{}^{\sigma\lambda}-\dfrac{1}{4}g^{\alpha\delta}R_{\nu\rho\sigma\lambda}R^{\beta\mu\nu\rho}R^{\eta}{}_{\mu}{}^{\sigma\lambda},
\end{align}
\begin{align}
    \mathcal{P}^{\alpha\beta\eta\delta}_{\beta_{11}}&=-4R_{\mu\nu\rho\sigma}R^{\mu\nu\rho\sigma}R^{\alpha\beta\delta\eta},\\
    \mathcal{P}^{\alpha\beta\eta\delta}_{\beta_{12}}&=-4R_{\mu\nu\rho\sigma}R^{\alpha\beta\mu\nu}R^{\delta\eta\rho\sigma},\\
    \mathcal{P}^{\alpha\beta\eta\delta}_{\beta_{13}}&=2R_{\mu\rho\nu\sigma}R^{\alpha\mu\eta\nu}R^{\beta\rho\delta\sigma}-2R_{\mu\rho\nu\sigma}R^{\alpha\mu\delta\nu}R^{\beta\rho\eta\sigma},\\
    \mathcal{P}^{\alpha\beta\eta\delta}_{\beta_{14}}&=-g^{\beta \eta}R^{\alpha\delta}R_{\mu\nu}R^{\mu\nu}+g^{\beta\delta}R^{\alpha\eta}R_{\mu\nu}R^{\mu\nu}+g^{\alpha\eta}R^{\beta\delta}R_{\mu\nu}R^{\mu\nu}-g^{\alpha\delta}R^{\beta\eta}R_{\mu\nu}R^{\mu\nu}, \\
\mathcal{P}^{\alpha\beta\eta\delta}_{\beta_{15}}&=\dfrac{1}{2}g^{\alpha\eta}g^{\beta\delta}R^{\mu\nu}R_{\mu}{}^{\rho\sigma\lambda}R_{\nu\rho\sigma\lambda}-\dfrac{1}{2}g^{\alpha\delta}g^{\beta\eta}R^{\mu\nu}R_{\mu}{}^{\rho\sigma\lambda}R_{\nu\rho\sigma\lambda}-\dfrac{1}{2}R^{\beta\mu}RR^{\alpha}{}_{\mu}{}^{\delta\eta}\nonumber\\
     &-\dfrac{1}{2}R^{\eta\mu}RR^{\alpha\beta\delta}{}_{\mu}+\dfrac{1}{2}R^{\delta\mu}RR^{\alpha\beta\eta}{}_{\mu}+\dfrac{1}{2}R^{\alpha\mu}RR^{\beta}{}_{\mu}{}^{\delta\eta}-\dfrac{1}{4}g^{\beta\eta}RR^{\alpha\mu\nu\rho}R^{\delta}{}_{\mu\nu\rho}\nonumber\\
     &+\dfrac{1}{4}g^{\alpha\eta}RR^{\beta\mu\nu\rho}R^{\delta}{}_{\mu\nu\rho}+\dfrac{1}{4}g^{\beta\delta}RR^{\alpha\mu\nu\rho}R^{\eta}{}_{\mu\nu\rho}-\dfrac{1}{4}g^{\alpha\delta}RR^{\beta\mu\nu\rho}R^{\eta}{}_{\mu\nu\rho}, \\
    \mathcal{P}^{\alpha\beta\eta\delta}_{\beta_{16}}&=-\dfrac{1}{2}R_{\mu}{}^{\nu}R^{\beta\mu}R^{\alpha}{}_{\nu}{}^{\delta\eta}-\dfrac{1}{4}g^{\beta\eta}R^{\delta\mu}R_{\mu\nu\rho\sigma}R^{\alpha\nu\rho\sigma}+\dfrac{1}{4}g^{\beta\delta}R^{\eta\mu}R_{\mu\nu\rho\sigma}R^{\alpha\nu\rho\sigma}-\dfrac{1}{2}R_{\mu}^{\nu}R^{\eta\mu}R^{\alpha\beta\delta}{}_{\nu}\nonumber\\
    &+\dfrac{1}{2}R_{\mu}^{\nu}R^{\delta\mu}R^{\alpha\beta\eta}{}_{\nu}+\dfrac{1}{2}R_{\mu}^{\nu}R^{\alpha\mu}R^{\beta}{}_{\nu}{}^{\delta\eta}+\dfrac{1}{4}g^{\alpha\eta}R^{\delta\mu}R_{\mu\nu\rho\sigma}R^{\beta\nu\rho\sigma}-\dfrac{1}{4}g^{\alpha\delta}R^{\eta\mu}R_{\mu\nu\rho\sigma}R^{\beta\nu\rho\sigma}\nonumber\\
    &-\dfrac{1}{4}g^{\beta\eta}R^{\alpha\mu}R_{\mu\nu\rho\sigma}R^{\delta\nu\rho\sigma}+\dfrac{1}{4}g^{\alpha\eta}R^{\beta\mu}R_{\mu\nu\rho\sigma}R^{\delta\nu\rho\sigma}+\dfrac{1}{4}g^{\beta\delta}R^{\alpha\mu}R_{\mu\nu\rho\sigma}R^{\eta\nu\rho\sigma}\nonumber\\
    &-\dfrac{1}{4}g^{\alpha\delta}R^{\beta\mu}R_{\mu\nu\rho\sigma}R^{\eta\nu\rho\sigma}, \\
\mathcal{P}^{\alpha\beta\eta\delta}_{\beta_{17}}&=\dfrac{1}{2}g^{\alpha\eta}g^{\beta\delta}R_{\mu\nu}{}^{\lambda\zeta}R^{\mu\nu\rho\sigma}R_{\rho\sigma\lambda\zeta}-\dfrac{1}{2}g^{\alpha\delta}g^{\beta\eta}R_{\mu\nu}{}^{\lambda\zeta}R^{\mu\nu\rho\sigma}R_{\rho\sigma\lambda\zeta}-3RR^{\alpha\beta\mu\nu}R^{\delta\eta}{}_{\mu\nu},\\ \mathcal{P}^{\alpha\beta\eta\delta}_{\beta_{18}}&=\dfrac{1}{2}g^{\alpha\eta}g^{\beta\delta}R_{\mu}{}^{\lambda}{}_{\rho}{}^{\zeta}R^{\mu\nu\rho\sigma}R_{\nu\lambda\sigma\zeta}-\dfrac{1}{2}g^{\alpha\delta}g^{\beta\eta}R_{\mu}{}^{\lambda}{}_{\rho}{}^{\zeta}R^{\mu\nu\rho\sigma}R_{\nu\lambda\sigma\zeta}+\dfrac{3}{2}RR^{\alpha\mu\eta\nu}R^{\beta}{}_{\mu}{}^{\delta}{}_{\nu}\nonumber\\
    &-\dfrac{3}{2}RR^{\alpha\mu\delta\nu}R^{\beta}{}_{\mu}{}^{\eta}{}_{\nu},\\
\mathcal{P}^{\alpha\beta\eta\delta}_{\beta_{19}}&=R_{\mu\nu\rho\sigma}R^{\alpha\nu\rho\sigma}R^{\beta\mu\rho\sigma}-R_{\mu\nu\rho\sigma}R^{\beta\nu\rho\sigma}R^{\alpha\mu\rho\sigma}+R_{\mu\nu\rho\sigma}R^{\alpha\beta\eta\mu}R^{\delta\nu\rho\sigma}-R_{\mu\nu\rho\sigma}R^{\alpha\beta\delta\mu}R^{\eta\nu\rho\sigma},\\
    \mathcal{P}^{\alpha\beta\eta\delta}_{\beta_{20}}&=-2R^{\alpha}{}_{\mu}{}^{\rho\sigma}R^{\beta}{}_{\nu\rho\sigma}R^{\delta\eta\mu\nu}-2R^{\alpha\beta\mu\nu}R^{\delta}{}_{\mu}{}^{\rho\sigma}R^{\eta}{}_{\nu\rho\sigma}, \\
   \mathcal{P}^{\alpha\beta\eta\delta}_{\beta_{21}}&=-\dfrac{1}{2}R^{\beta\mu}R^{\eta\nu}R^{\alpha}{}_{\mu}{}^{\delta}{}_{\nu}+\dfrac{1}{2}R^{\beta\mu}R^{\delta\nu}R^{\alpha}{}_{\mu}{}^{\eta}{}_{\nu}+\dfrac{1}{2}R^{\alpha\mu}R^{\eta\nu}R^{\beta}{}_{\mu}{}^{\delta}{}_{\nu}-\dfrac{1}{2}R^{\alpha\mu}R^{\delta\nu}R^{\beta}{}_{\mu}{}^{\eta}{}_{\nu}\nonumber\\
   &-\dfrac{1}{2}g^{\beta\eta}R^{\mu\nu}R^{\alpha\rho}{}_{\mu}{}^{\sigma}R^{\delta}{}_{\rho\nu\sigma}+\dfrac{1}{2}g^{\alpha\eta}R^{\mu\nu}R^{\beta\rho}{}_{\mu}{}^{\sigma}R^{\delta}{}_{\rho\nu\sigma}+\dfrac{1}{2}g^{\beta\delta}R^{\mu\nu}R^{\alpha\rho}{}_{\mu}{}^{\sigma}R^{\eta}{}_{\rho\nu\sigma}\nonumber\\
   &-\dfrac{1}{2}g^{\alpha\delta}R^{\mu\nu}R^{\beta\rho}{}_{\mu}{}^{\sigma}R^{\eta}{}_{\rho\nu\sigma},\\
\mathcal{P}^{\alpha\beta\eta\delta}_{\beta_{22}}&=-R^{\alpha\rho}{}_{\mu}{}^{\sigma}R^{\beta\mu\eta\nu}R^{\delta}{}_{\rho\nu\sigma}+R^{\alpha\mu\eta\nu}R^{\beta\rho}{}_{\mu}{}^{\sigma}R^{\delta}{}_{\rho\nu\sigma}+R^{\alpha\rho}{}_{\mu}{}^{\sigma}R^{\beta\mu\delta\nu}R^{\eta}{}_{\rho\nu\sigma}-R^{\alpha\mu\delta\nu}R^{\beta\rho}{}_{\mu}{}^{\sigma}R^{\eta}{}_{\rho\nu\sigma},
\end{align}
\begin{align}
   \mathcal{P}^{\alpha\beta\eta\delta}_{\beta_{23}}&=-\dfrac{1}{4}g^{\beta\eta}R_{\mu}{}^{\rho\sigma\lambda}R_{\nu\rho\sigma\lambda}R^{\alpha\mu\delta\nu}+\dfrac{1}{4}g^{\beta\delta}R_{\mu}{}^{\rho\sigma\lambda}R_{\nu\rho\sigma\lambda}R^{\alpha\mu\eta\nu}+\dfrac{1}{2}R^{\mu\nu}R^{\alpha\rho\delta\eta}R^{\beta}{}_{\mu\nu\rho}\nonumber\\
   &+\dfrac{1}{4}g^{\alpha\eta}R_{\mu}{}^{\rho\sigma\lambda}R_{\nu\rho\sigma\lambda}R^{\beta\mu\delta\nu}-\dfrac{1}{4}g^{\alpha\delta}R_{\mu}{}^{\rho\sigma\lambda}R_{\nu\rho\sigma\lambda}R^{\beta\mu\eta\nu}-\dfrac{1}{2}R^{\mu\nu}R^{\alpha}{}_{\mu\nu\rho}R^{\beta\rho\delta\eta}\nonumber\\
   &-\dfrac{1}{4}R^{\beta\eta}R^{\alpha\mu\nu\rho}R^{\delta}{}_{\mu\nu\rho}-\dfrac{1}{2}R^{\mu\nu}R^{\alpha\beta\eta\rho}R^{\delta}{}_{\mu\nu\rho}+\dfrac{1}{4}R^{\alpha\eta}R^{\beta\mu\nu\rho}R^{\delta}{}_{\mu\nu\rho}+\dfrac{1}{4}R^{\beta\delta}R^{\alpha\mu\nu\rho}R^{\eta}{}_{\mu\nu\rho}\nonumber\\
   &+\dfrac{1}{2}R^{\mu\nu}R^{\alpha\beta\delta\rho}R^{\eta}{}_{\mu\nu\rho}-\dfrac{1}{4}R^{\alpha\delta}R^{\beta\mu\nu\rho}R^{\eta}{}_{\mu\nu\rho}, \\
   \mathcal{P}^{\alpha\beta\eta\delta}_{\beta_{24}}&=-\dfrac{1}{4}R^{\eta\mu}R^{\alpha\nu\delta\rho}R^{\beta}{}_{\nu\mu\rho}+\dfrac{1}{4}R^{\delta\mu}R^{\alpha\nu\eta\rho}R^{\beta}{}_{\nu\mu\rho}+\dfrac{1}{4}R^{\mu\nu}R^{\alpha}{}_{\mu}{}^{\eta\rho}R^{\beta}{}_{\nu}{}^{\delta}{}_{\rho}\nonumber\\
   &-\dfrac{1}{4}R^{\mu\nu}R^{\alpha}{}_{\mu}{}^{\delta\rho}R^{\beta}{}_{\nu}{}^{\eta}{}_{\rho}+\dfrac{1}{4}R^{\eta\mu}R^{\alpha}{}_{\nu\mu\rho}R^{\beta\nu\delta\rho}-\dfrac{1}{4}R^{\delta\mu}R^{\alpha}{}_{\nu\mu\rho}R^{\beta\nu\eta\rho}\nonumber\\
   &+\dfrac{1}{4}R^{\mu\nu}R^{\alpha\rho\eta}{}_{\mu}R^{\beta}{}_{\rho}{}^{\delta}{}_{\nu}-\dfrac{1}{4}R^{\mu\nu}R^{\alpha\rho\delta}{}_{\mu}R^{\beta}{}_{\rho}{}^{\eta}{}_{\nu}+\dfrac{1}{4}R^{\beta\mu}R^{\alpha\nu\eta\rho}R^{\delta}{}_{\rho\mu\nu}\nonumber\\
   &-\dfrac{1}{4}R^{\alpha\mu}R^{\beta\nu\eta\rho}R^{\delta}{}_{\rho\mu\nu}-\dfrac{1}{4}g^{\beta\eta}R_{\mu\sigma\rho\lambda}R^{\alpha\mu\nu\rho}R^{\delta\sigma}{}_{\nu}{}^{\lambda}+\dfrac{1}{4}g^{\alpha\eta}R_{\mu\sigma\rho\lambda}R^{\beta\mu\nu\rho}R^{\delta\sigma}{}_{\nu}{}^{\lambda}\nonumber\\
   &-\dfrac{1}{4}R^{\beta\mu}R^{\alpha\nu\delta\rho}R^{\eta}{}_{\rho\mu\nu}+\dfrac{1}{4}R^{\alpha\mu}R^{\beta\nu\delta\rho}R^{\eta}{}_{\rho\mu\nu}+\dfrac{1}{4}g^{\beta\delta}R_{\mu\sigma\rho\lambda}R^{\alpha\mu\nu\rho}R^{\eta\sigma}{}_{\nu}{}^{\lambda}\nonumber\\
   &-\dfrac{1}{4}g^{\alpha\delta}R_{\mu\sigma\rho\lambda}R^{\beta\mu\nu\rho}R^{\eta\sigma}{}_{\nu}{}^{\lambda}.
\end{align}

\section{Symplectic Potential and Noether Wald charges for $\mathcal{R}^4$ terms}\label{expressionsquartic}
Let us first report on the {symplectic potential} for the 26 quartic invariants as parameterized in \eqref{lagquart}. Consider that $\Theta_{\alpha_{0}}=\Theta_{\beta_{0}}$, given in \eqref{fofRTheta}. From where we see the usefulness of parameterizing powers of the Ricci scalar in the general function $f_0(R)$. For the remaining terms we have
{\small
\begin{align}
    \Theta^{m}_{\beta_{1}}&=-\nabla_{a}hR^{am}R^{2}+h\nabla_{b}\left(R^{2}R^{m b}\right)+2R^{ab}R^{2}\nabla_{b}h^{m}_{a}-2h_{a}^{c}\nabla_{c}\left(R^{2}R^{am}\right)-\nabla^{m}h_{ab}R^{ab}R^{2}\nonumber\\
    &+h_{ab}\nabla^{m}\left(R^{ab}R^{2}\right)+2\nabla_{c}h^{cm}RR_{ab}R^{ab}-2h^{m d}\nabla_{d}\left(RR_{ab}R^{ab}\right)-2\nabla^{m}hRR_{ab}R^{ab}\nonumber\\
    &+2h\nabla^{m}\left(RR_{ab}R^{ab}\right),\\
    \Theta^{m}_{\beta_{2}}&=-\dfrac{3}{2}\nabla_{b}hRR^{ab}R_{a}^{m}+\dfrac{3}{2}h\nabla_{c}\left(RR^{am}R_{a}^{c}\right)+3\nabla_{c}h_{b}^{m}RR^{ab}R_{a}^{c}-3h_{b}^{d}\nabla_{d}\left(RR^{ab}R_{a}^{m}\right)\nonumber\\
    &-\dfrac{3}{2}\nabla^{m}h_{bc}RR^{ab}R_{a}^{c}+\dfrac{3}{2}h_{bc}\nabla_{m}\left(RR^{ab}R_{a}^{c}\right)+\nabla_{d}h^{dm}R_{a}^{c}R^{ab}R_{bc}-h^{em}\nabla_{e}\left(R_{a}^{c}R^{ab}R_{bc}\right)\nonumber\\
    &-\nabla^{m}h R_{a}^{c}R^{ab}R_{bc}+h\nabla^{m}\left(R_{a}^{c}R^{ab}R_{bc}\right),\\
     \Theta^{m}_{\beta_{3}}&=-2\nabla_{c}hR_{a}^{c}R^{ab}R_{b}^{m}+2h\nabla_{d}\left(R_{a}^{m}R^{ab}R_{b}^{d}\right)+4\nabla_{d}h_{c}^{m}R_{a}^{c}R^{ab}R_{b}^{d}-4h_{c}^{e}\nabla_{e}\left(R_{a}^{c}R^{ab}R_{b}^{m}\right)\nonumber\\
     &-2\nabla^{m}h_{cd}R_{a}^{c}R^{ab}R_{b}^{d}+2h_{cd}\nabla^{m}\left(R_{a}^{c}R^{ab}R_{b}^{d}\right),\\     
     \Theta^{m}_{\beta_{4}}&=\nabla_{b}h_{ac}RR^{ab}R^{cm}-h_{ac}\nabla_{d}\left(RR^{am}R
     ^{cd}\right)-\nabla_{c}h_{ab}RR^{cm}R^{ab}+h_{ab}\nabla_{d}\left(RR^{dm}R^{ab}\right)\nonumber\\
     &-\nabla^{m}h^{cd}RR^{ab}R_{acbd}+h^{cd}\nabla^{m}\left(RR^{ab}R_{acbd}\right)+2\nabla_{c}h^{cd}RR^{ab}R_{adb}{}^{m}-2h^{dm}\nabla^{e}\left(RR^{ab}R_{adbe}\right)\nonumber\\
     &-\nabla^{d}hRR^{ab}R_{adb}{}^{m}+h\nabla^{e}\left(RR^{ab}R_{a}{}^{m}{}_{be}\right)+\nabla_{e}h^{em}R^{ab}R^{cd}R_{acbd}-h^{fm}\nabla_{f}\left(R^{ab}R^{cd}R_{acbd}\right)\nonumber\\
     &-\nabla^{m}hR^{ab}R^{cd}R_{acbd}+h\nabla^{m}\left(R^{ab}R^{cd}R_{acbd}\right), \\
\Theta^{m}_{\beta_{5}}&=-\dfrac{1}{2}\nabla_{b}h_{de}R^{m}_{a}R^{ab}R^{de}+\dfrac{1}{2}h_{de}\nabla_{c}\left(R_{a}^{c}R^{am}R^{de}\right)-\nabla_{a}h^{fm}R^{ab}R^{cd}R_{bcdf}+h^{ef}\nabla_{e}\left(R^{bm}R^{cd}R_{bcdf}\right)\nonumber\\
     &+\nabla_{c}h_{bd}R_{a}^{c}R^{ab}R^{m d}-h_{bd}\nabla_{e}\left(R_{a}^{m}R^{ab}R^{de}\right)-\dfrac{1}{2}\nabla_{d}h_{bc}R_{a}^{c}R^{ab}R^{dm}+\dfrac{1}{2}h_{bc}\nabla_{e}\left(R_{a}^{c}R^{ab}R^{em}\right)\nonumber\\
     &+\nabla^{m}h_{a}^{e}R^{ab}R^{cd}R_{bcde}-h_{a}^{e}\nabla^{m}\left(R^{ab}R^{cd}R_{bcde}\right)-\dfrac{1}{2}\nabla^{m}h^{de}R_{a}^{c}R^{ab}R_{bdce}+\dfrac{1}{2}h^{de}\nabla^{m}\left(R_{a}^{c}R^{ab}R_{bdce}\right)\nonumber\\
     &+\nabla_{a}hR^{ab}R^{cd}R_{bcd}{}^{m}-h\nabla^{f}\left(R^{bm}R^{cd}R_{bcdf}\right)+\nabla_{d}h^{de}R_{a}^{c}R^{ab}R_{bec}{}^{m}-h^{em}\nabla^{f}\left(R_{a}^{c}R^{ab}R_{becf}\right)\nonumber\\
     &-\nabla_{e}h_{a}^{e}R^{ab}R^{cd}R_{bcd}{}^{m}+h_{a}^{m}\nabla^{f}\left(R^{ab}R^{cd}R_{bcdf}\right)-\dfrac{1}{2}\nabla^{e}hR_{a}^{c}R^{ab}R_{bec}{}^{m}+\dfrac{1}{2}h\nabla^{f}\left(R_{a}^{c}R^{ab}R_{b}{}^{m}{}_{cf}\right), \\
\Theta^{m}_{\beta_{6}}&= 2\nabla_{a}h^{am} RR_{cdef}R^{cdef}-2h^{bm}\nabla_{b}\left(R R_{cdef}R^{cdef}\right)-2\nabla^{m}hRR_{cdef}R^{cdef}\nonumber\\
&+2h\nabla^{m}\left(RR_{cdef}R^{cdef}\right)-4\nabla^{c}h^{ab}R^{2}R_{acb}{}^{m}+4h^{ab}\nabla^{d}\left(R^2R_{a}{}^{m}{}_{bd}\right),\\
\Theta^{m}_{\beta_{7}}&=-\nabla_{a}h R^{am}R_{defj}R^{defj}+h\nabla_{b}\left(R^{bm}R_{defj}R^{defj}\right)+2\nabla_{b}h_{a}^{m}R^{ab}R_{defj}R^{defj}\nonumber\\
 &-2h_{a}^{c}\nabla_{c}\left(R^{am}R_{defj}R^{defj}\right)-\nabla^{m}h_{ab}R^{ab}R_{defj}R^{defj}+h_{ab}\nabla^{m}\left(R^{ab}R_{defj}R^{defj}\right)\nonumber\\
 &-4\nabla^{e}h^{cd}R_{ab}R^{ab}R_{ced}{}^{m}+4h^{cd}\nabla^{f}\left(R_{ab}R^{ab}R_{c}{}^{m}{}_{df}\right),
\end{align}
\begin{align}
\Theta^{m}_{\beta_{8}}&= -\nabla^{m}h^{cd}R^{ab}R_{ac}{}^{fj}R_{bdfj}+h^{cd}\nabla^{m}\left(R^{ab}R_{ac}{}^{fj}R_{bdfj}\right)+2\nabla_{c}h^{cd}R^{ab}R_{ad}{}^{fj}R_{b}{}^{m}{}_{fj}\nonumber\\
&-2h^{dm}\nabla^{e}\left(R^{ab}R_{ad}{}^{fj}R_{befj}\right)-\nabla^{d}hR^{ab}R_{ad}{}^{fj}R_{b}{}^{m}{}_{fj}+h\nabla^{e}\left(R^{ab}R_{a}{}^{m fj}R_{befj}\right)\nonumber\\
&-4\nabla_{c}h_{a}^{e}R^{ab}R^{cd}R_{bde}{}^{m}+4h_{a}^{e}\nabla^{f}\left(R^{ab}R^{dm}R_{bdef}\right), \\
\Theta^{m}_{\beta_{9}}&=-\nabla_{a}h^{ef}R^{am}R^{cd}R_{cedf}+h^{ef}\nabla_{b}\left(R^{m b}R^{cd}R_{cedf}\right)-\nabla^{m}h^{cd}R^{ab}R_{a}{}^{f}{}_{b}{}^{j}R_{cfdj}\nonumber\\
&+h^{cd}\nabla^{m}\left(R^{ab}R_{a}{}^{f}{}_{b}{}^{j}R_{cfdj}\right)+2\nabla_{c}h^{cd}R^{ab}R_{a}{}^{f}{}_{b}{}^{j}R_{df}{}^{m}{}_{j}-2h^{m d}\nabla^{e}\left(R^{ab}R_{a}{}^{f}{}_{b}{}^{j}R_{dfej}\right)\nonumber\\
&-\nabla^{d}hR^{ab}R_{a}{}^{f}{}_{b}{}^{j}R_{df}{}^{m}{}_{j}+h\nabla^{e}\left(R^{ab}R_{a}{}^{f}{}_{b}{}^{j}R^{m}{}_{fej}\right)+2\nabla_{b}h_{a}^{e}R^{ab}R^{cd}R_{ced}{}^{m}\nonumber\\
&-2h_{a}^{e}\nabla^{f}\left(R^{am}R^{cd}R_{cedf}\right)-\nabla^{e}h_{ab}R^{ab}R^{cd}R_{ced}{}^{m}+h_{ab}\nabla^{f}\left(R^{ab}R^{cd}R_{c}{}^{m}{}_{df}\right),\\
\Theta^{m}_{\beta_{10}}&= -\dfrac{1}{2}\nabla^{m}h^{ab}R_{a}{}^{def}R_{bd}{}^{jk}R_{efjk}+\dfrac{1}{2}h^{ab}\nabla^{m}\left(R_{a}{}^{def}R_{bd}{}^{jk}R_{efjk}\right)+\nabla_{a}h^{ab}R_{b}{}^{def}R^{m}{}_{d}{}^{jk}R_{efjk}\nonumber\\
  &-h^{m b}\nabla^{c}\left(R_{b}{}^{def}R_{cd}{}^{jk}R_{efjk}\right)-\dfrac{1}{2}\nabla^{b}hR_{b}{}^{def}R^{m}{}_{d}{}^{jk}R_{efjk}+\dfrac{1}{2}h\nabla^{c}\left(R^{m def}R_{cd}{}^{jk}R_{efjk}\right)\nonumber\\
  &+2\nabla_{a}h^{cd}R^{ab}R_{bc}{}^{fj}R_{d}{}^{m}{}_{fj}-2h^{cd}\nabla^{e}\left(R^{m b}R_{bc}{}^{fj}R_{defj}\right)-2\nabla^{d}h_{a}^{c}R^{ab}R_{bd}{}^{fj}R_{c}{}^{m}{}_{fj}\nonumber\\
  &+2h_{a}^{c}\nabla^{e}\left(R^{ab}R_{b}{}^{m fj}R_{cefj}\right)-2\nabla^{d}h_{a}^{c}R^{ab}R_{bc}{}^{fj}R_{d}{}^{m}{}_{fj}+2h_{a}^{c}\nabla^{e}\left(R^{ab}R_{bc}{}^{fj}R^{m}{}_{efj}\right)\nonumber\\
  &-2\nabla^{e}h^{cd}R^{ab}R_{a}{}^{j}{}_{ce}R_{bjd}{}^{m}+2h^{cd}\nabla^{f}\left(R^{ab}R_{a}{}^{j}{}_{c}{}^{m}R_{bjdf}\right), \\ \Theta^{m}_{\beta_{11}}&=-8\nabla^{c}h^{ab}R_{acb}{}^{m}R_{efjk}R^{efjk}+8h^{ab}\nabla^{d}\left(R_{a}{}^{m}{}_{bd}R_{efjk}R^{efjk}\right),\\
  \Theta^{m}_{\beta_{12}}&=-8\nabla^{c}h^{ab}R_{ac}{}^{ef}R_{b}{}^{m jk}R_{efjk}+8h^{ab}\nabla^{d}\left(R_{a}{}^{m ef}R_{bd}{}^{jk}R_{efjk}\right),\\
  \Theta^{m}_{\beta_{13}}&=-4\nabla^{c}h^{ab}R_{a}{}^{e}{}_{b}{}^{f}R_{c}{}^{jm k}R_{ejfk}+4h^{ab}\nabla^{d}\left(R_{a}{}^{e}{}_{b}{}^{f}R^{m j}{}_{d}{}^{k}R_{ejfk}\right)+ 4\nabla^{c}h^{ab}R_{a}{}^{e}{}_{c}{}^{f}R_{b}{}^{jm k}R_{ekfj}\nonumber\\
  &-4h^{ab}\nabla^{d}\left(R_{a}{}^{em f}R_{b}{}^{j}{}_{d}{}^{k}R_{ekfj}\right),  \\
\Theta^{m}_{\beta_{14}}&=-2\nabla_{c}hR_{ab}R^{ab}R^{cm}+
2h\nabla_{d}\left(R_{ab}R^{ab}R^{dm}\right)+4\nabla_{d}h_{c}^{m}R_{ab}R^{ab}R^{cd}-
4h_{c}^{e}\nabla_{e}\left(R_{ab}R^{ab}R^{cm}\right)\nonumber\\
&-2\nabla^{m}h_{cd}R_{ab}R^{ab}R^{cd}+2h_{cd}\nabla^{m}
\left(R_{ab}R^{ab}R^{cd}\right),\\
\Theta^{m}_{\beta_{15}}&=-\dfrac{1}{2}\nabla^{m}h^{ab}RR_{a}{}^{def}R_{bdef}+\dfrac{1}{2}h^{ab}\nabla^{m}\left(RR_{a}{}^{def}R_{bdef}\right)+\nabla_{a}h^{ab}RR_{b}{}^{def}R^{m}{}_{def}\nonumber\\
&-h^{m a}\nabla^{c}\left(RR_{a}{}^{def}R_{cdef}\right)-\dfrac{1}{2}\nabla^{b}hRR_{b}{}^{def}R^{m}{}_{def}+\dfrac{1}{2}h\nabla^{c}\left(RR^{m def}R_{cdef}\right)\nonumber\\
&+\nabla_{c}h^{cm}R^{ab}R_{a}{}^{efj}R_{befj}-h^{dm}\nabla_{d}\left(R^{ab}R_{a}{}^{efj}R_{befj}\right)-\nabla^{m}hR^{ab}R_{a}{}^{efj}R_{befj}\nonumber\\
&+h\nabla^{m}\left(R^{ab}R_{a}{}^{efj}R_{befj}\right)+2\nabla_{a}h^{cd}RR^{ab}R_{bcd}{}^{m}-2h^{cd}\nabla^{e}\left(RR^{m b}R_{bcde}\right)\nonumber\\
&-2\nabla^{d}h_{a}^{c}RR^{ab}R_{bcd}{}^{m}+2h_{a}^{c}\nabla^{e}\left(RR^{ab}R_{bc}{}^{m}{}_{e}\right)-2\nabla^{d}h_{a}^{c}RR^{ab}R_{bdc}{}^{m}+2h_{a}^{c}\nabla^{e}\left(RR^{ab}R_{b}{}^{m}{}_{ce}\right), 
\end{align}
\begin{align}
\Theta^{m}_{\beta_{16}}&=\nabla_{a}h^{m d}R^{ab}R_{b}{}^{efj}R_{defj}-h^{cd}\nabla_{c}\left(R^{m b}R_{b}{}^{efj}R_{defj}\right)-\nabla^{m}h_{a}^{c}R^{ab}R_{b}{}^{efj}R_{cefj}\nonumber\\
&+h_{a}^{c}\nabla^{m}\left(R^{ab}R_{b}{}^{efj}R_{cefj}\right)-\nabla_{a}hR^{ab}R_{b}{}^{efj}R^{m}{}_{efj}+h\nabla^{d}\left(R^{m b}R_{b}{}^{efj}R_{defj}\right)\nonumber\\
&+\nabla_{c}h_{a}^{c}R^{ab}R_{b}{}^{efj}R^{m}{}_{efj}-h_{a}^{m}\nabla^{d}\left(R^{ab}R_{b}{}^{efj}R_{defj}\right)+2\nabla_{b}h^{de}R_{a}^{c}R^{ab}R_{cde}{}^{m}\nonumber\\
&-2h^{de}\nabla^{f}\left(R_{a}^{c}R^{am}R_{cdef}\right)-2\nabla^{e}
h_{b}^{d}R_{a}^{c}R^{ab}R_{cde}{}^{m}+2h_{b}^{d}\nabla^{f}\left(R_{a}^{c}R^{ab}R_{cd}{}^{m}{}_{f}\right)\nonumber\\
&-2\nabla^{e}h_{b}^{d}R_{a}^{c}R^{ab}R_{ced}{}^{m}+2h_{b}^{d}\nabla^{f}\left(R_{a}^{c}R^{ab}R_{c}{}^{m}{}_{df}\right),\\
\Theta^{m}_{\beta_{17}}&=\nabla_{a}h^{am}R_{cd}{}^{jk}R^{cdef}R_{efjk}-h^{bm}\nabla_{b}\left(R_{cd}{}^{jk}R^{cdef}R_{efjk}\right)-\nabla^{m}hR_{cd}{}^{jk}R^{cdef}R_{efjk}\nonumber\\
&+h\nabla^{m}\left(R_{cd}{}^{jk}R^{cdef}R_{efjk}\right)-6\nabla^{c}h^{ab}RR_{ac}{}^{ef}R_{b}{}^{m}{}_{ef}+6h^{ab}\nabla^{d}\left(RR_{a}{}^{m ef}R_{bdef}\right), \\
\Theta^{m}_{\beta_{18}}&=\nabla_{a}h^{am}R_{c}{}^{j}{}_{e}{}^{k}R^{cdef}R_{djfk}-h^{bm}\nabla_{b}\left(R_{c}{}^{j}{}_{e}{}^{k}R^{cdef}R_{djfk}\right)-\nabla^{m}hR_{c}{}^{j}{}_{e}{}^{k}R^{cdef}R_{djfk}\nonumber\\
&+h\nabla^{m}\left(R_{c}{}^{j}{}_{e}{}^{k}R^{cdef}R_{djfk}\right)+3\nabla^{c}h^{ab}RR_{a}{}^{e}{}_{c}{}^{f}R_{bf}{}^{m}{}_{e}-3h^{ab}\nabla^{d}\left(RR_{a}{}^{em f}R_{bfde}\right)\nonumber\\
&-3\nabla^{c}h^{ab}RR_{a}{}^{e}{}_{b}{}^{f}R_{cd}{}^{m}{}_{f}+3h^{ab}\nabla^{d}\left(RR_{a}{}^{e}{}_{b}{}^{f}R^{m}{}_{edf}\right), \\
\Theta^{m}_{\beta_{19}}&=4\nabla^{c}h^{ab}R_{a}{}^{m}{}_{c}{}^{e}R_{b}{}^{fjk}R_{efjk}-4h^{ab}\nabla^{d}\left(R_{ad}{}^{m e}R_{b}{}^{fjk}R_{efjk}\right)+4\nabla^{c}h^{ab}R_{a}{}^{e}{}_{c}{}^{m}R_{b}{}^{fjk}R_{efjk}\nonumber\\
&-4h^{ab}\nabla^{d}\left(R_{a}{}^{em}{}_{d}R_{b}{}^{fjk}R_{efjk}\right)-4\nabla^{c}h^{ab}R_{a}{}^{m}{}_{b}{}^{e}R_{c}{}^{fjk}R_{efjk}+4h^{ab}\nabla^{d}\left(R_{adb}{}^{e}R^{m fjk}R_{efjk}\right),\\
\Theta^{m}_{\beta_{20}}&=-4\nabla^{c}h^{ab}R_{a}{}^{m ef}R_{be}{}^{jk}R_{cfjk}
+4h^{ab}\nabla^{d}\left(R_{ad}{}^{ef}R_{be}{}^{jk}R^{m}{}_{fjk}\right)-4\nabla^{c}h^{ab}R_{ac}{}^{ef}R_{be}{}^{jk}R^{m}{}_{fjk}\nonumber\\
&+4h^{ab}\nabla^{d}\left(R_{a}{}^{m ef}R_{be}{}^{jk}R_{dfjk}\right), \\
\Theta^{m}_{\beta_{21}}&=-\nabla_{a}h^{ef}R^{ab}R^{m d}R_{bedf}+h^{ef}\nabla_{c}\left(R^{m b}R^{cd}R_{bedf}\right)-\nabla^{m}h^{cd}R^{ab}R_{a}{}^{f}{}_{c}{}^{j}R_{bfdj}\nonumber\\
    &+h^{cd}\nabla^{m}\left(R^{ab}R_{a}{}^{f}{}_{c}{}^{j}R_{bfdj}\right)+2\nabla_{c}h^{cd}R^{ab}R_{a}{}^{f}{}_{d}{}^{j}R_{bf}{}^{m}{}_{j}-2h^{m d}\nabla^{e}\left(R^{ab}R_{a}{}^{f}{}_{d}{}^{j}R_{bfej}\right)\nonumber\\
    &-\nabla^{d}hR^{ab}R_{a}{}^{f}{}_{d}{}^{j}R_{bf}{}^{m}{}_{j}+h\nabla^{e}\left(R^{ab}R_{a}{}^{fm j}R_{bfej}\right)+2\nabla_{c}h_{a}^{e}R^{ab}R^{cd}R_{b}{}^{m}{}_{de}\nonumber\\
    &-2h_{a}^{e}\nabla^{f}\left(R^{ab}R^{m d}R_{bfde}\right)-\nabla^{e}h_{ac}R^{ab}R^{cd}R_{bed}{}^{m}+h_{ac}\nabla^{f}\left(R^{ab}R^{cd}R_{b}{}^{m}{}_{df}\right), \\
\Theta^{m}_{\beta_{22}}&=-2\nabla^{c}h^{ab}R_{a}{}^{efj}R_{bef}{}^{k}R_{cj}{}^{m}{}_{k}+2h^{ab}\nabla^{d}\left(R_{a}{}^{efj}R_{bef}{}^{k}R^{m}{}_{jdk}\right)+2\nabla^{c}h^{ab}R_{a}{}^{em f}R_{b}{}^{j}{}_{f}{}^{k}R_{cjek}\nonumber\\
    &-2h^{ab}\nabla^{d}\left(R_{a}{}^{e}{}_{d}{}^{f}R_{b}{}^{j}{}_{f}{}^{k}R^{m}{}_{jek}\right)+2\nabla^{c}h^{ab}R_{a}{}^{e}{}_{c}{}^{f}R_{b}{}^{j}{}_{f}{}^{k}R^{m}{}_{jek}-2h^{ab}\nabla^{d}\left(R_{a}{}^{em f}R_{b}{}^{j}{}_{f}{}^{k}R_{djek}\right)\nonumber\\
    &-2\nabla^{c}h^{ab}R_{a}{}^{e}{}_{b}{}^{f}R_{c}{}^{j}{}_{e}{}^{k}R^{m}{}_{jfk}+2h^{ab}\nabla^{d}\left(R_{a}{}^{e}{}_{b}{}^{f}R^{m j}{}_{e}{}^{k}R_{djfk}\right), 
\end{align}
\begin{align}
{\Theta^{m}_{\beta_{23}}}&={-\dfrac{1}{2}\nabla_{a}h^{cd}R^{am}R_{c}{}^{efj}R_{defj}+\dfrac{1}{2}h^{cd}\nabla_{b}\left(R^{bm}R_{c}{}^{efj}R_{defj}\right)-\dfrac{1}{2}\nabla^{m}h^{ab}R_{a}{}^{d}{}_{b}{}^{e}R_{d}{}^{fjk}R_{efjk}}\nonumber\\
    &+\dfrac{1}{2}h^{ab}\nabla^{m}\left(R_{a}{}^{d}{}_{b}{}^{e}R_{d}{}^{fjk}R_{efjk}\right)+\nabla_{a}h^{ab}R_{b}{}^{dm e}R_{d}{}^{fjk}R_{efjk}-h^{bm}\nabla^{c}\left(R_{b}{}^{d}{}_{c}{}^{e}R_{d}{}^{fjk}R_{efjk}\right)\nonumber\\
    &-\dfrac{1}{2}\nabla^{b}hR_{b}{}^{dm e}R_{d}{}^{fjk}R_{efjk}+\dfrac{1}{2}h\nabla^{c}\left(R^{m d}{}_{c}{}^{e}R_{d}{}^{fjk}R_{efjk}\right)+\nabla_{b}h_{a}^{c}R^{ab}R_{c}{}^{efj}R^{m}{}_{efj}\nonumber\\
    &-h_{a}^{c}\nabla^{d}\left(R^{am}R_{c}{}^{efj}R_{defj}\right)-\dfrac{1}{2}\nabla^{c}h_{ab}R^{ab}R_{c}{}^{efj}R^{m}{}_{efj}+\dfrac{1}{2}h_{ab}\nabla^{d}\left(R^{ab}R^{m efj}R_{defj}\right)\nonumber\\
    &-2\nabla^{e}h^{cd}R^{ab}R_{aeb}{}^{j}R_{c}{}^{m}{}_{dj}+2h^{cd}\nabla^{f}\left(R^{ab}R_{a}{}^{m}{}_{b}{}^{j}R_{cfdj}\right)+2\nabla^{e}h^{cd}R^{ab}R_{acb}{}^{j}R_{d}{}^{m}{}_{ej}\nonumber\\
    &-2h^{cd}\nabla^{f}\left(R^{ab}R_{acb}{}^{j}R_{df}{}^{m}{}_{j}\right)+2\nabla^{e}h^{cd}R^{ab}R_{acb}{}^{j}R_{dje}{}^{m}-2h^{cd}\nabla^{f}\left(R^{ab}R_{acb}{}^{j}R_{dj}{}^{m}{}_{f}\right), \\
\Theta^{m}_{\beta_{24}}&=-\dfrac{1}{2}\nabla^{m}h^{ab}R_{a}{}^{def}R_{b}{}^{j}{}_{e}{}^{k}R_{djfk}+\dfrac{1}{2}h^{ab}\nabla^{m}\left(R_{a}{}^{def}R_{b}{}^{j}{}_{e}{}^{k}R_{djfk}\right)+\nabla_{a}h^{ab}R_{b}{}^{def}R^{m j}{}_{e}{}^{k}R_{djfk}\nonumber\\
    &-h^{m b}\nabla^{c}\left(R_{b}{}^{def}R_{c}{}^{j}{}_{e}{}^{k}R_{djfk}\right)-\dfrac{1}{2}\nabla^{b}h R_{b}{}^{def}R^{m j}{}_{e}{}^{k}R_{djfk}+\dfrac{1}{2}h\nabla^{c}\left(R^{m def}R_{c}{}^{j}{}_{e}{}^{k}R_{djfk}\right)\nonumber\\
    &-\nabla_{a}h^{cd}R^{ab}R_{b}{}^{fm j}R_{cfdj}+h^{cd}\nabla^{e}\left(R^{m b}R_{b}{}^{f}{}_{e}{}^{j}R_{cfdj}\right)+\nabla_{a}h^{cd}R^{ab}R_{b}{}^{f}{}_{c}{}^{j}R_{df}{}^{m}{}_{j}\nonumber\\
    &-h^{cd}\nabla^{e}\left(R^{bm}R_{b}{}^{f}{}_{c}{}^{j}R_{dfej}\right)+\nabla^{d}h_{a}^{c}R^{ab}R_{b}{}^{fm j}R_{cfdj}-h_{a}^{c}\nabla^{e}\left(R^{ab}R_{b}{}^{f}{}_{e}{}^{j}R_{cf}{}^{m}{}_{j}\right)\nonumber\\
    &-\nabla^{d}h_{a}^{c}R^{ab}R_{b}{}^{f}{}_{d}{}^{j}R_{cf}{}^{m}{}_{j}+h_{a}^{c}\nabla^{e}\left(R^{ab}R_{b}{}^{fm j}R_{cfej}\right)+\nabla^{d}h_{a}^{c}R^{ab}R_{b}{}^{fm j}R_{cjdf}\nonumber\\
    &-h_{a}^{c}\nabla^{e}\left(R^{ab}R_{b}{}^{f}{}_{e}{}^{j}R_{cj}{}^{m}{}_{f}\right)-\nabla^{d}h_{a}^{c}R^{ab}R_{b}{}^{f}{}_{c}{}^{j}R_{df}{}^{m}{}_{j}+h_{a}^{c}\nabla^{e}\left(R^{ab}R_{b}{}^{f}{}_{c}{}^{j}R^{m}{}_{fej}\right)\nonumber\\
    &+\nabla^{e}h^{cd}R^{ab}R_{ac}{}^{m j}R_{bedj}-h^{cd}\nabla^{f}\left(R^{ab}R_{acf}{}^{j}R_{b}{}^{m}{}_{dj}\right)-\nabla^{e}h^{cd}R^{ab}R_{acd}{}^{j}R_{be}{}^{m}{}_{j}\nonumber\\
    &+h^{cd}\nabla^{f}\left(R^{ab}R_{acd}{}^{j}R_{b}{}^{m}{}_{fj}\right), \\
\Theta^{m}_{\beta_{25}}&= -\nabla^{c}h^{ab}R_{a}{}^{efj}R_{b}{}^{k}{}_{fj}R_{ce}{}^{m}{}_{k}+h^{ab}\nabla^{d}\left(R_{a}{}^{efj}R_{b}{}^{k}{}_{fj}R^{m}{}_{edk}\right)+\nabla^{c}h^{ab}R_{a}{}^{e m f}R_{bf}{}^{jk}R_{cejk}\nonumber\\
   &-h^{ab}\nabla^{d}\left(R_{a}{}^{e}{}_{d}{}^{f}R_{bf}{}^{jk}R^{m}{}_{ejk}\right)+\nabla^{c}h^{ab}R_{a}{}^{emf}R_{be}{}^{jk}R_{cfjk}-h^{ab}\nabla^{d}\left(R_{a}{}^{e}{}_{d}{}^{f}R_{be}{}^{jk}R^{m}{}_{fjk}\right)\nonumber\\
   &-4\nabla^{c}h^{ab}R_{a}{}^{mef}R_{b}{}^{j}{}_{e}{}^{k}R_{cjfk}+4h^{ab}\nabla^{d}\left(R_{ad}{}^{ef}R_{b}{}^{j}{}_{e}{}^{k}R^{m}{}_{jfk}\right)+\nabla^{c}h^{ab}R_{a}{}^{e}{}_{c}{}^{f}R_{bf}{}^{jk}R^{m}{}_{ejk}\nonumber\\
   &-h^{ab}\nabla^{d}\left(R_{a}{}^{emf}R_{bf}{}^{jk}R_{dejk}\right)-\nabla^{c}h^{ab}R_{a}{}^{e}{}_{c}{}^{f}R_{be}{}^{jk}R^{m}{}_{fjk}+h^{ab}\nabla^{d}\left(R_{a}{}^{emf}R_{be}{}^{jk}R_{dfjk}\right)\nonumber\\
   &-\nabla^{c}h^{ab}R_{a}{}^{e}{}_{b}{}^{f}R_{ce}{}^{jk}R^{m}{}_{fjk}+h^{ab}\nabla^{d}\left(R_{a}{}^{e}{}_{b}{}^{f}R^{m}{}_{e}{}^{jk}R_{dfjk}\right).
\end{align}
}
Now, in order to provide the final details of the expressions that would permit a direct implementation on a symbolic algebra package for tensor calculus, we provide the {Noether-Wald charge} for each of the quartic terms, as parameterized in \eqref{lagquart}. Of course, the contribution of $\beta_0$, can be read from the contribution of $\alpha_0$, as given in \eqref{NWalpha0}. For the remaining terms, we have the lengthy formulae
{\small
\begin{align}
    Q^{mn}_{\beta_{1}}&=4\xi^{[n}\nabla_{b}\left(R^{m]b}R^{2}\right)+4\nabla_{b}\xi^{[m}R^{n]b}R^{2}+4\xi_{b}\nabla^{[m}\left(R^{n]b}R^{2}\right)+4\nabla^{[n}\xi^{m]}RR_{ab}R^{ab}\nonumber\\
    &+8\xi^{[n}\nabla^{m]}\left(RR_{ab}R^{ab}\right),\\
    Q^{mn}_{\beta_{2}}&=6\nabla_{b}\xi^{[m}R^{n]a}RR_{a}^{b}+6\xi^{[n}\nabla_{b}\left(R^{m]a}RR_{a}^{b}\right)+6\xi_{e}\nabla^{[m}\left(R^{n]a}RR_{a}^{e}\right)+2\nabla^{[n}\xi^{m]}R_{a}^{c}R^{ab}R_{bc}\nonumber\\
    &+4\xi^{[n}\nabla^{m]}\left(R_{a}^{c}R^{ab}R_{bc}\right),\\
    Q^{mn}_{\beta_{3}}&=8\nabla_{d}\xi^{[m}R^{n]}_{a}R^{ab}R_{b}^{d}-8\xi^{[m}\nabla_{d}\left(R_{b}^{n]}R^{ab}R_{a}^{d}\right)+8\xi_{d}\nabla^{[m}\left(R^{n]}_{a}R^{ab}R_{b}^{d}\right),\\
    Q^{mn}_{\beta_{4}}&=2\nabla_{b}\xi_{c}RR^{b[n}R^{m]c}+4\xi_{c}\nabla_{a}\left(RR^{a[m}R^{n]c}\right)+4\nabla^{[n}\xi^{d}R_{adb}{}^{m]}RR^{ab}+4\xi^{d}\nabla^{[m}\left(R_{adb}{}^{n]}RR^{ab}\right)\nonumber\\
    &+4\xi^{[n}\nabla^{e}\left(R_{a}{}^{m]}{}_{be}RR^{ab}\right)+2\nabla^{[n}\xi^{m]}R^{ab}R^{cd}R_{acbd}+4\xi^{[n}\nabla^{m]}\left(R^{ab}R^{cd}R_{acbd}\right),\\
       Q^{mn}_{\beta_{5}}&=2\nabla_{b}\xi_{d}R_{a}^{b}R^{a[n}R^{m]d}+2\xi_{d}\nabla_{e}\left(R_{a}^{e}R^{a[m}R^{n]d}\right)+2\xi_{d}\nabla_{e}\left(R_{a}^{[n}R^{m]e}R^{ad}\right)+2\nabla_{a}\xi^{[n}R_{bcd}{}^{m]}R^{ab}R^{cd}\nonumber\\
    &+2\nabla^{[m}\xi_{f}R_{bcd}{}^{n]}R^{fb}R^{cd}+2\nabla^{[n}\xi^{f}R_{bfc}{}^{m]}R_{a}^{c}R^{ab}+4\xi^{[m}\nabla^{f}\left(R^{n]b}R^{cd}R_{bcdf}\right)\nonumber\\
    &+2\xi^{[n}\nabla^{f}\left(R_{b}{}^{m]}{}_{cf}R_{a}^{c}R^{ab}\right)+2\xi^{f}\nabla^{[n}\left(R^{m]b}R^{cd}R_{bcdf}\right)+2\xi_{a}\nabla^{[n}\left(R^{ab}R^{cd}R_{bcd}{}^{m]}\right)\nonumber\\
    &+2\xi^{e}\nabla_{e}\left(R^{cd}R^{b[m}R_{bcd}{}^{n]}\right)+2\xi^{d}\nabla^{[m}\left(R_{a}^{c}R^{ab}R_{bdc}{}^{n]}\right), \\
Q^{mn}_{\beta_{6}}&=4\nabla^{[n}\xi^{m]}RR_{cdef}R^{cdef}+8\xi^{[n}\nabla^{m]}\left(RR_{cdef}R^{cdef}\right)+8\nabla^{a}\xi^{b}R^{2}R_{b}{}^{[m}{}_{a}{}^{n]}+8\xi^{b}\nabla^{c}\left(R^{2}R_{bc}{}^{nm}\right)\nonumber\\
    &+8\xi^{b}\nabla^{a}\left(R^2R_{a}{}^{[n}{}_{b}{}^{m]}\right),\\
Q^{mn}_{\beta_{7}}&=4\nabla_{b}\xi^{[m}R^{n]b}R_{defj}R^{defj}+4\xi^{[n}\nabla_{b}\left(R^{m]b}R_{defj}R^{defj}\right)+4\xi_{b}\nabla^{[m}\left(R^{n]b}R_{defj}R^{defj}\right)\nonumber\\
    &+8\nabla^{d}\xi^{e}R_{ab}R^{ab}R_{e}{}^{[m}{}_{d}{}^{n]}+8\xi^{d}\nabla^{f}\left(R_{ab}R^{ab}R_{df}{}^{nm}\right),\\
    Q^{mn}_{\beta_{8}}&=4\nabla^{[n}\xi^{d}R^{ab}R_{ad}{}^{fj}R_{b}{}^{m]}{}_{fj}-8\nabla_{c}\xi_{a}R^{ab}R^{cd}R_{bd}{}^{[nm]}+4\xi^{d}\nabla^{[m}\left(R^{ab}R_{a}{}^{n]fj}R_{bdfj}\right)\nonumber\\
    &+4\xi^{[n}\nabla^{a}\left(R^{eb}R_{e}{}^{m]fj}R_{bafj}\right)+4\xi^{d}\nabla^{a}\left(R^{b[n}R^{m]f}R_{bfda}\right)+8\xi_{d}\nabla^{a}\left(R^{db}R^{f[n}R_{bfa}{}^{m]}\right)\nonumber\\
    &+8\nabla^{c}\left(R_{e}^{b}R^{a[m}R_{abc}{}^{n]}\right)\xi^{e}-8R_{e}^{b}R^{a[m}R_{ab}{}^{n]}{}_{d}\nabla^{d}\xi^{e},\\
    Q^{mn}_{\beta_{9}}&=4\nabla_{a}\xi^{f}R^{a[n}R_{cfd}{}^{m]}R^{cd}+4\xi^{f}\nabla_{b}\left(R^{b[m}R_{c}{}^{n]}{}_{df}R^{cd}\right)+4\xi_{f}\nabla^{b}\left(R^{f[n}R_{c}{}^{m]}{}_{db}R^{cd}\right)\nonumber\\
    &+4\nabla^{[n}\xi^{d}R_{dj}{}^{m]}{}_{f}R^{ab}R_{a}{}^{f}{}_{b}{}^{j}+4\xi^{d}\nabla^{[m}\left(R^{n]}{}_{fdj}R^{ab}R_{a}{}^{f}{}_{b}{}^{j}\right)+4\xi^{[n}\nabla^{d}\left(R^{m]}{}_{fdj}R^{ab}R_{a}{}^{f}{}_{b}{}^{j}\right), \\
Q^{mn}_{\beta_{10}}&=2\nabla^{[n}\xi^{b}R^{m]}{}_{d}{}^{jk}R_{b}{}^{def}R_{efjk}+2\xi^{b}\nabla^{[m}\left(R^{n]def}R_{bd}{}^{jk}R_{efjk}\right)+2\xi^{[n}\nabla^{c}\left(R^{m]def}R_{cdjk}R_{ef}{}^{jk}\right)\nonumber\\
    &-4\nabla^{c}\xi_{d}R^{db}R_{bc}{}^{fj}R^{[nm]}{}_{fj}+4\nabla^{c}\xi^{d}R^{ab}R_{a}{}^{j}{}_{c}{}^{[n}R_{bjd}{}^{m]}+4\xi^{d}\nabla^{a}\left(R^{b[n}R_{b}{}^{m]fj}R_{dafj}\right)\nonumber\\
    &+4\xi^{d}\nabla^{a}\left(R^{b[n}R^{m]}{}_{afj}R_{bd}{}^{fj}\right)+4\xi_{d}\nabla^{a}\left(R^{db}R_{b}{}^{[m fj}R^{n]}{}_{afj}\right)+4\xi^{d}\nabla^{a}\left(R^{fb}R_{f}{}^{j}{}_{da}R_{bj}{}^{[nm]}\right),
  \end{align}    
\begin{align}
Q^{mn}_{\beta_{11}}&=16R_{efjk}R^{efjk}\nabla^{b}\xi^{a}R_{a}{}^{[m}{}_{b}{}^{n]}+16\xi^{b}\nabla^{a}\left(R_{ba}{}^{mn}R_{efjk}R^{efjk}\right),\\
     Q^{mn}_{\beta_{12}}&=16\nabla^{b}\xi^{a}R_{efjk}R^{jk}{}_{a}{}^{[m}R_{b}{}^{n]ef}+16\xi^{b}\nabla^{a}\left(R_{efjk}R^{ef}{}_{a}{}^{[m}R_{b}{}^{n]jk}+R_{efjk}R_{ba}{}^{ef}R^{mn jk}\right),\\
     Q^{mn}_{\beta_{13}}&=8\nabla^{c}\xi^{b}R^{[n e}{}_{c}{}^{f}R^{m] k}{}_{b}{}^{j}R_{ekfj}+8\xi^{b}\nabla_{a}\left(R_{b}{}^{eaf}R_{ejfk}R^{[n\; j\; m] k}\right)-8\xi^{b}\nabla^{a}\left(R_{b}{}^{ef[n}R^{m]j}{}_{a}{}^{k}R_{ejfk}\right)\nonumber\\
     &-8\xi^{b}\nabla^{a}\left(R_{b}{}^{ef[n}R^{m]j}{}_{a}{}^{k}R_{fjek}\right), \\
Q^{mn}_{\beta_{14}}&=8\nabla_{d}\xi^{[m}R^{n]d}R_{ab}R^{ab}+8\xi^{[n}\nabla_{d}
\left(R^{m]d}R_{ab}R^{ab}\right)+8\xi_{d}\nabla^{[m}\left(R^{n]d}R_{ab}R^{ab}\right),\\
Q^{mn}_{\beta_{15}}&=-4\nabla^{c}\xi_{d}RR^{db}R_{bc}{}^{[nm]}+4\xi^{c}\nabla^{e}\left(R^{b[n}R_{bc}{}^{m]}{}_{e}\right)+4\xi^{c}\nabla^{e}\left(RR^{b[n}R_{b}{}^{m]}{}_{ce}\right)+4\xi_{c}\nabla^{e}\left(RR^{cb}R_{b}{}^{[n}{}_{e}{}^{m]}\right)\nonumber\\
&+2\nabla^{[n}\xi^{b}R^{m]}{}_{def}RR_{b}{}^{def}+2\nabla^{[n}\xi^{m]}R^{ab}R_{a}{}^{efj}R_{befj}+4\xi^{[n}\nabla^{m]}\left(R^{ab}R_{a}{}^{efj}R_{befj}\right)\nonumber\\
&+2\xi^{b}\nabla^{[m}\left(R^{n]def}RR_{bdef}\right)+2\xi^{[n}\nabla^{c}\left(R^{m]def}R_{cdef}R\right), \\
Q^{mn}_{\beta_{16}}&=-4\nabla^{d}\xi_{b}R_{a}^{c}R^{ab}R_{cd}{}^{[nm]}+4\xi_{d}\nabla^{f}\left(R_{a}^{c}R^{ad}R_{c}{}^{[mn]}{}_{f}\right)+4\xi^{d}\nabla^{f}\left(R_{a}^{c}R^{a[n}R^{m]}{}_{fcd}\right)\nonumber\\
&+4\xi^{d}\nabla^{f}\left(R_{a}^{c}R^{a[n}R_{c}{}^{m]}{}_{df}\right)+2\nabla_{a}\xi^{[m}R^{n]}{}_{efj}R^{ab}R_{b}{}^{efj}+2\nabla^{[n}\xi_{a}R^{m]}{}_{efj}R^{ab}R_{b}{}^{efj}\nonumber\\
&+2\xi^{d}\nabla_{d}\left(R^{b[n}R^{m]}{}_{efj}R_{b}{}^{efj}\right)+4\xi^{[n}\nabla^{d}\left(R^{m]b}R_{b}{}^{efj}R_{defj}\right)+2\xi_{d}\nabla^{[m}\left(R^{n]}{}_{efj}R^{db}R_{b}{}^{efj}\right)\nonumber\\
&+2\xi^{d}\nabla^{[m}\left(R^{n]b}R_{b}{}^{efj}R_{defj}\right),\\
Q^{mn}_{\beta_{17}}&=12\nabla^{b}\xi^{a}RR_{b}{}^{[n ef}R_{a}{}^{m]}{}_{ef}+2\nabla^{[n}\xi^{m]}R_{cd}{}^{jk}R^{cdef}R_{efjk}+4\xi^{[n}\nabla^{m]}\left(R_{cd}{}^{jk}R^{cdef}R_{efjk}\right)\nonumber\\
&+12\xi^{b}\nabla^{a}\left(RR_{ba}{}^{ef}R^{[nm]}{}_{ef}\right),\\
Q^{mn}_{\beta_{18}}&=2\nabla^{[n}\xi^{m]}R_{c}{}^{j}{}_{e}{}^{k}R^{cdef}R_{djfk}+4\xi^{[n}\nabla^{m]}\left(R_{c}{}^{j}{}_{e}{}^{k}R^{cdef}R_{djfk}\right)+6\nabla^{b}\xi^{a}RR_{a}{}^{e}{}_{b}{}^{f}R^{[n}{}_{f}{}^{m]}{}_{e}\nonumber\\
&+6\nabla^{a}\xi^{b}RR_{a}{}^{e[n f}R_{bf}{}^{m]}{}_{e}+6\xi^{a}\nabla^{d}\left(RR^{[n e}{}_{a}{}^{f}R^{m]}{}_{edf}\right)+6\xi^{a}\nabla^{d}\left(RR_{a}{}^{e[n f}R_{de}{}^{m]}{}_{f}\right), \\
Q^{mn}_{\beta_{19}}&=8\nabla^{c}\xi^{b}R^{[nm]}{}_{c}{}^{e}R_{b}{}^{fjk}R_{efjk}+8\xi^{b}\nabla^{a}\left(R_{ba}{}^{[n e}R^{m]fjk}R_{efjk}\right)+8\xi^{b}\nabla^{a}\left(R^{[n}{}_{ab}{}^{e}R^{m]fjk}R_{efjk}\right)\nonumber\\
&+8\xi^{b}\nabla^{a}\left(R_{a}{}^{[nm]e}R_{b}{}^{fjk}R_{efjk}\right),\\
Q^{mn}_{\beta_{20}}&=8\nabla^{a}\xi^{b}R_{a}{}^{[n e f}R^{m]}{}_{fjk}R_{be}{}^{jk}+8\nabla^{b}\xi^{a}R_{a}{}^{[n e f}R^{m]}{}_{e}{}^{jk}R_{bfjk}+8\xi^{b}\nabla^{a}\left(R^{[nm]ef}R_{be}{}^{jk}R_{afjk}\right)\nonumber\\
&+8\xi^{b}\nabla^{a}\left(R_{ba}{}^{ef}R^{[m}{}_{fjk}R^{n]}{}_{e}{}^{jk}\right), \\
Q^{mn}_{\beta_{21}}&=4\nabla_{a}\xi^{f}R^{ab}R^{d[n}R_{b}{}^{m]}{}_{df}+4\nabla_{a}\xi_{f}R^{fb}R^{ad}R_{b}{}^{[m}{}_{d}{}^{n]}+4\nabla^{[n}\xi^{d}R^{ab}R_{b}{}^{f}{}_{d}{}^{j}R_{af}{}^{m]}{}_{j}\nonumber\\
&+4\xi^{d}\nabla^{[m}\left(R^{ab}R_{a}{}^{fn] j}R_{bfdj}\right)+4\xi^{[n}\nabla^{e}\left(R^{ab}R_{a}{}^{fm]j}R_{bfej}\right)+4\xi^{e}\nabla_{a}\left(R^{b[m}R^{a d}R_{b}{}^{n]}{}_{de}\right)\nonumber\\
&+2\xi^{e}\nabla^{a}\left(R^{b[n}R^{m]d}R_{be da}\right)+4\xi_{e}\nabla^{a}\left(R^{b[n}R^{e d}R_{b}{}^{m]}{}_{d a}\right),\\
Q^{mn}_{\beta_{22}}&=4\nabla^{a}\xi^{b}R^{[m}{}_{e\;f\;j}R_{a}{}^{j\;n]\; k}R_{b}{}^{e}{}^{f}{}_{k}+4\nabla^{a}\xi^{b}R_{b}{}^{e[m f}R^{n] j}{}_{f}{}^{k}R_{a\;j\;e\;k}+4\nabla^{a}\xi^{b}R_{b}{}^{e}{}^{f}{}_{j}R_{a\;e\;f\;k}R^{[n\;k\;m]j},
\end{align}
\begin{align}
    Q^{mn}_{\beta_{23}}&=2\nabla_{a}\xi^{d}R^{a[n}R^{m]}{}_{efj}R_{d}{}^{efj}+2\nabla^{[n}\xi^{b}R_{b}{}^{dm]e}R_{d}{}^{fjk}R_{efjk}+4\nabla^{a}\xi^{d}R^{eb}R_{edb}{}^{j}R_{aj}{}^{[nm]}\nonumber\\
    &+2\xi^{[n}\nabla^{c}\left(R^{m]d}{}_{c}{}^{e}R_{d}{}^{fjk}R_{efjk}\right)+2\xi^{b}\nabla^{[m}\left(R^{n]d}{}_{b}{}^{e}R_{d}{}^{fjk}R_{efjk}\right)+2\xi^{d}\nabla^{c}\left(R_{c}{}^{[m}R^{n]efj}R_{defj}\right)\nonumber\\
    &+2\xi_{d}\nabla^{c}\left(R^{d[n}R^{m]efj}R_{cefj}\right)+4\xi^{d}\nabla^{c}\left(R^{ab}R_{a}{}^{[m}{}_{b}{}^{j}R^{n]}{}_{cdj}\right)+4\xi^{d}\nabla^{c}\left(R^{ab}R_{a}{}^{[m}{}_{b}{}^{j}R_{dc}{}^{n]}{}_{j}\right)\nonumber\\
    &+4\xi^{d}\nabla^{c}\left(R^{ab}R_{adb}{}^{j}R^{[m}{}_{c}{}^{n]}{}_{j}\right), \\
   Q^{mn}_{\beta_{24}}&=2\nabla^{[n}\xi^{b}R^{m]j}{}_{e}{}^{k}R_{b}{}^{def}R_{djfk}+4\nabla^{c}\xi^{d}R_{d}^{b}R_{b}{}^{f[m j}R_{cf}{}^{n]}{}_{j}+2\nabla^{c}\xi^{d}R^{ab}R_{ad}{}^{[m j}R_{bc}{}^{n]}{}_{j}+\nonumber\\
   &+2\xi^{a}\nabla^{[m}\left(R^{n]def}R_{a}{}^{j}{}_{e}{}^{k}R_{djfk}\right)+2\xi^{d}\nabla^{a}\left(R^{b[n}R_{b}{}^{fm]j}R_{dfaj}\right)+2\xi^{d}\nabla^{a}\left(R^{b[m}R_{b}{}^{f}{}_{a}{}^{j}R_{df}{}^{n]}{}_{j}\right)\nonumber\\
   &+2\xi^{d}\nabla^{a}\left(R^{b[m}R_{b}{}^{f}{}_{a}{}^{j}R^{n]}{}_{fdj}\right)+2\xi^{d}\nabla^{a}\nabla^{a}\left(R^{b[n}R_{b}{}^{f}{}_{d}{}^{j}R^{m]}{}_{faj}\right)+2\xi^{[n}\nabla^{c}\left(R^{m]def}R_{c}{}^{j}{}_{e}{}^{k}R_{djfk}\right)\nonumber\\
   &+2\xi_{d}\nabla^{a}\left(R^{db}R_{b}{}^{f[n j}R^{m]}{}_{faj}\right)+2\xi^{d}\nabla^{a}\left(R^{eb}R_{eda}{}^{j}R_{b}{}^{[nm]}{}_{j}\right)+2\xi_{d}\nabla^{a}\left(R^{db}R_{b}{}^{f}{}_{a}{}^{j}R^{[n}{}_{f}{}^{m]}{}_{j}\right)\nonumber\\
   &+2\xi^{d}\nabla^{a}\left(R^{eb}R_{ed}{}^{[n j}R_{b}{}^{m]}{}_{aj}\right)+2\xi^{d}\nabla^{a}\left(R^{eb}R_{e}{}^{[n}{}_{d}{}^{j}R_{ba}{}^{m]}{}_{j}\right)+2\xi^{d}\nabla^{a}\left(R^{eb}R_{e}{}^{[n}{}_{d}{}^{j}R_{b}{}^{m]}{}_{aj}\right), \\
Q^{mn}_{\beta_{25}}&=4\nabla^{a}\xi^{b}R_{b}{}^{e[m f}R^{n]}{}_{f}{}^{jk}R_{aejk}+2\nabla^{a}\xi^{b}R_{a}{}^{e[m f}R^{n]}{}_{fjk}R_{be}{}^{jk}-2\nabla^{a}\xi^{b}R_{b}{}^{efj}R^{[n k}{}_{fj}R_{ae}{}^{m]}{}_{k}.
\end{align}
}

\end{document}